\documentclass{aa}

\usepackage{psfig,natbib}

\begin{document}

\newcommand{\MHI}{\mbox{$M_{\mbox{\tiny HI}}$}}
\newcommand{\NHI}{\mbox{$N_{\mbox{\tiny HI}}$}}
\newcommand{\VHI}{\mbox{$V_{\mbox{\tiny HI}}$}}
\newcommand{\Vopt}{\mbox{$V_{\mbox{\tiny opt}}$}}
\newcommand{\NH}{\mbox{$N_{\mbox{\tiny H}}$}}
\newcommand{\Lb}{\mbox{$L_{\mbox{\tiny B}}$}}
\newcommand{\LbA}{\mbox{$L_{\mbox{\tiny B}}^{\mbox{\tiny A}}$}}
\newcommand{\LbB}{\mbox{$L_{\mbox{\tiny B}}^{\mbox{\tiny B}}$}}
\newcommand{\Lv}{\mbox{$L_{\mbox{\tiny V}}$}}
\newcommand{\Lr}{\mbox{$L_{\mbox{\tiny R}}$}}
\newcommand{\LHa}{\mbox{$L_{\mbox{\tiny H}\alpha}$}}
\newcommand{\Ha}{\mbox{H$_{\alpha}$}}
\newcommand{\Hb}{\mbox{H$_{\beta}$}}
\newcommand{\Hg}{\mbox{H$_{\gamma}$}}
\newcommand{\Hd}{\mbox{H$_{\delta}$}}
\newcommand{\Mb}{\mbox{$M_{\mbox{\tiny B}}$}}
\newcommand{\Mv}{\mbox{$M_{\mbox{\tiny V}}$}}
\newcommand{\Mr}{\mbox{$M_{\mbox{\tiny R}}$}}
\newcommand{\MHH}{\mbox{$M_{\mbox{\tiny H$_2$}}$}}
\newcommand{\HH}{\mbox{H$_2$}}
\newcommand{\Mo}{\mbox{M$_{\odot}$}}
\newcommand{\Mvir}{\mbox{$M_{\mbox{\tiny vir}}$}}
\newcommand{\Mdyn}{\mbox{$M_{\mbox{\tiny dyn}}$}}
\newcommand{\Mmar}{\mbox{$M_{\mbox{\tiny mar}}$}}
\newcommand{\Lo}{\mbox{L$_{\odot}$}}
\newcommand{\Zo}{\mbox{Z$_{\odot}$}}
\newcommand{\x}{\mbox{$\times$}}
\newcommand{\Dph}{\mbox{$D_{25}$}}
\newcommand{\Vrot}{\mbox{$V_{\mbox{\tiny rot}}$}}
\newcommand{\rH}{\mbox{$r_{\mbox{\tiny H}}$}}
\newcommand{\Mk}{\mbox{$M_{\mbox{\tiny K}}$}}
\newcommand{\COa}{\mbox{CO(1$\rightarrow$0)}}
\newcommand{\COb}{\mbox{CO(1$\rightarrow$1)}}
\newcommand{\CO}{\mbox{CO}}
\newcommand{\cmm}{\mbox{cm$^{-2}$}}
\newcommand{\cmmm}{\mbox{cm$^{-3}$}}
\newcommand{\Mpc}{\mbox{Mpc}}
\newcommand{\kms}{\mbox{km~s$^{-1}$}}
\newcommand{\usfr}{\mbox{M$_{\odot}$~yr$^{-1}$}}
\newcommand{\pcc}{\mbox{pc$^{-2}$}}
\newcommand{\kpc}{\mbox{kpc}}
\newcommand{\sbr}{\mbox{mag/\fbox{}\arcsec}}
\newcommand{\sbB}{\mbox{$\mu _{\mbox{\tiny B}}$}}
\newcommand{\sbV}{\mbox{$\mu _{\mbox{\tiny V}}$}}
\newcommand{\sbc}{\mbox{$\mu _{\mbox{\tiny B}0}$}}
\newcommand{\fB}{\mbox{$f_{\mbox{\tiny B}}$}}
\newcommand{\aop}{\mbox{$a_{25}$}}
\newcommand{\bop}{\mbox{$b_{25}$}}
\newcommand{\tabsp}{\noalign{\smallskip}}
\newcommand{\BV}{\mbox{$\mbox{B} - \mbox{V}$}}
\newcommand{\VK}{\mbox{$\mbox{V} - \mbox{K}$}}
\newcommand{\sigint}{\mbox{$\sigma _{\mbox{\tiny int}}$}}
\newcommand{\Hoa}{\mbox{H$_0= 75$~km~s$^{-1}$~Mpc$^{-1}$}}
\newcommand{\Hob}{\mbox{H$_0= 70$~km~s$^{-1}$~Mpc$^{-1}$}}
\newcommand{\qo}{\mbox{q$_0$}}
\newcommand{\Mbo}{\mbox{$M_{\odot}^{\mbox{\tiny B}}$}}
\newcommand{\Vpar}{\mbox{$V_{\mbox{\tiny par}}$}}
\newcommand{\eqw}{\mbox{$W$(H$_{\beta}$)}}
\newcommand{\Ab}{\mbox{$A_{\mbox{\tiny B}}$}}
\newcommand{\Av}{\mbox{A$_{\mbox{\tiny V}}$}}
\newcommand{\noteb}{\mbox{$^{\mbox{\tiny +}}$}}
\newcommand{\mJyb}{\mbox{mJy~beam$^{-1}$}}
\newcommand{\mJy}{\mbox{mJy}}
\newcommand{\K}{\mbox{K}}
\newcommand{\Ts}{\mbox{$T_{\mbox{s}}$}}
\newcommand{\TOIII}{\mbox{$T_e({\mbox{OIII}}$)}}
\newcommand{\TOII}{\mbox{$T_e({\mbox{OIII}}$)}}
\newcommand{\micron}{\mbox{$\mu$m}}
\newcommand{\Oabun}{\mbox{$12 + \log(\frac{\mbox{O}}{\mbox{H}})$}}
\newcommand{\er}{\mbox{$\pm$}}

\newcommand{\fb}{\mbox{$f_{\mbox{\tiny B}}$}}
\newcommand{\Lfir}{\mbox{$L_{\mbox{\tiny FIR}}$}}
\newcommand{\Lir}{\mbox{$L_{\mbox{\tiny IR}}$}}
\newcommand{\uflux}{\mbox{erg~cm$^{-2}$~s$^{-1}$}}
\newcommand{\usbflux}{\mbox{erg~s$^{-1}$~cm$^{-2}$~arcsec$^{-2}$}}
\newcommand{\ufluxm}{\mbox{erg~cm$^{-2}$~s$^{-1}$~\AA$^{-1}$}}
\newcommand{\ul}{\mbox{erg~s$^{-1}$}}
\newcommand{\uSFR}{\mbox{M$_{\odot}$~yr$^{-1}$}}
\newcommand{\muJy}{\mbox{$\mu$Jy}}
\newcommand{\OIIIa}{\mbox{[OIII]$_{\lambda 4959}$}}
\newcommand{\OIIIb}{\mbox{[OIII]$_{\lambda 5007}$}}
\newcommand{\OIIIc}{\mbox{[OIII]$_{\lambda 4363}$}}
\newcommand{\OI}{\mbox{[OI]$_{\lambda 6300}$}}
\newcommand{\OII}{\mbox{[OII]$_{\lambda 3727}$}}
\newcommand{\OIIb}{\mbox{[OII]$_{\lambda 7320,30}$}}

\newcommand{\OIIIt}{\mbox{[OIII]}}
\newcommand{\OIIt}{\mbox{[OII]}}
\newcommand{\OIt}{\mbox{[OI]}}
\newcommand{\SIIt}{\mbox{[SII]}}
\newcommand{\NIIt}{\mbox{[NII]}}
\newcommand{\NIt}{\mbox{[NI]}}
\newcommand{\ArIIIt}{\mbox{[ArIII]}}

\newcommand{\NIIa}{\mbox{[NII]$_{\lambda 6548}$}}
\newcommand{\NIIb}{\mbox{[NII]$_{\lambda 6584}$}}
\newcommand{\SIIa}{\mbox{[SII]$_{\lambda 6717}$}}
\newcommand{\SIIb}{\mbox{[SII]$_{\lambda 6731}$}}
\newcommand{\SII}{\mbox{[SII]$_{\lambda 6717,6731}$}}

\newcommand\cola {\null}
\newcommand\colb {&}
\newcommand\colc {&}
\newcommand\cold {&}
\newcommand\cole {&}
\newcommand\colf {&}
\newcommand\colg {&}
\newcommand\colh {&}
\newcommand\coli {&}
\newcommand\colj {&}
\newcommand\colk {&}
\newcommand\coll {&}
\newcommand\colm {&}
\newcommand\coln {&}
\newcommand\eol{\\}
\newcommand\extline{&&&&&&&&&\eol}

% ------------- Additional definitions by Polis -----------------------------
%
\def\HI{H\,{\small I}}
\def\HI{H\,{\sc i}}
\def\HII{H\,{\sc ii}}
\def\HIit{\mbox{H\hspace{0.155 em}{\footnotesize \it I}}}
\def\nan{Nan\c{c}ay}
\def\sbu{${\rm mag\,\,arcsec^{-2 }} $ \ }
\newcommand{\am}[2]{$#1'\,\hspace{-1.7mm}.\hspace{.0mm}#2$}
\newcommand{\as}[2]{$#1''\,\hspace{-1.7mm}.\hspace{.0mm}#2$}
\newcommand\btab[5]{\begin{table*}[#1]{\parbox{#4}{\caption{#2}}\rule[-0.5ex]{0cm}{0.5ex} }
\begin{tabular*}{#4}{#5} }
% SMALL Letters
\newcommand\sbtab[5]{\begin{table}[#1]{\parbox{#4}{\caption{#2}}\rule[-0.5ex]{0cm}{0.5ex} }
%\label{#3}
\begin{footnotesize}
\begin{tabular*}{#4}{#5} }
\newcommand{\etab}[4]{
\end{tabular*}
\vspace*{#1}
\begin{flushleft}
\parbox{#2}{#3}
\end{flushleft}
\label{#4}
\end{table*} }
\def\kato{\rule[-1.25ex]{0cm}{1.25ex}}
\def\pano{\rule[0.0ex]{0cm}{2.5ex}}
% ------------------------------------------------------------------
%
% Additional abbreviations for journals
%
%
% ------------------------------------------------------------------
%
%  These Macros are taken from the AAS TeX macro package version 4.0.
%  Include this file in your LaTeX source only if you are not using
%  the AAS TeX macro package and need to resolve the macro definitions
%  in the BibTeX entries returned by the ADS abstract service.
%
%  For more information on the AASTeX macro package, please see the URL
%	http://www.ferberts.com/AAS/aastex.html
%  For more information about ADS abstract server, please see the URL
%	http://adswww.harvard.edu/ads_abstracts.html
%

% Abbreviations for journals.  The object here is to provide authors
% with convenient shorthands for the most "popular" (often-cited)
% journals; the author can use these markup tags without being concerned
% about the exact form of the journal abbreviation, or its formatting.
% It is up to the keeper of the macros to make sure the macros expand
% to the proper text.  If macro package writers agree to all use the
% same TeX command name, authors only have to remember one thing, and
% the style file will take care of editorial preferences.  This also
% applies when a single journal decides to revamp its abbreviating
% scheme, as happened with the ApJ (Abt 1991).

% \let\jnl@style=\rm
% \def\rf@jnl#1{{\jnl@style#1}}
\def\rf@jnl#1{{#1}}
\def\aj{\rf@jnl{AJ }}                   % Astronomical Journal
\def\araa{\rf@jnl{ARA\&A }}             % Annual Review of Astron and Astrophys
\def\apj{\rf@jnl{ApJ }}                 % Astrophysical Journal
\def\apjl{\rf@jnl{ApJ }}                % Astrophysical Journal, Letters
\def\apjs{\rf@jnl{ApJS }}               % Astrophysical Journal, Supplement
\def\ao{\rf@jnl{Appl.~Opt.}}           % Applied Optics
\def\apss{\rf@jnl{Ap\&SS }}             % Astrophysics and Space Science
\def\aap{\rf@jnl{A\&A }}                % Astronomy and Astrophysics
\def\aapr{\rf@jnl{A\&A~Rev.}}          % Astronomy and Astrophysics Reviews
\def\aaps{\rf@jnl{A\&AS }}              % Astronomy and Astrophysics, Supplement
\def\azh{\rf@jnl{AZh }}                 % Astronomicheskii Zhurnal
\def\baas{\rf@jnl{BAAS }}               % Bulletin of the AAS
\def\jrasc{\rf@jnl{JRASC }}             % Journal of the RAS of Canada
\def\memras{\rf@jnl{MmRAS }}            % Memoirs of the RAS
\def\mnras{\rf@jnl{MNRAS }}             % Monthly Notices of the RAS
\def\pra{\rf@jnl{Phys.~Rev.~A}}        % Physical Review A: General Physics
\def\prb{\rf@jnl{Phys.~Rev.~B}}        % Physical Review B: Solid State
\def\prc{\rf@jnl{Phys.~Rev.~C}}        % Physical Review C
\def\prd{\rf@jnl{Phys.~Rev.~D}}        % Physical Review D
\def\pre{\rf@jnl{Phys.~Rev.~E}}        % Physical Review E
\def\prl{\rf@jnl{Phys.~Rev.~Lett.}}    % Physical Review Letters
\def\pasp{\rf@jnl{PASP }}               % Publications of the ASP
\def\pasj{\rf@jnl{PASJ }}               % Publications of the ASJ
\def\qjras{\rf@jnl{QJRAS }}             % Quarterly Journal of the RAS
\def\skytel{\rf@jnl{S\&T }}             % Sky and Telescope
\def\solphys{\rf@jnl{Sol.~Phys.}}      % Solar Physics
\def\sovast{\rf@jnl{Soviet~Ast.}}      % Soviet Astronomy
\def\ssr{\rf@jnl{Space~Sci.~Rev.}}     % Space Science Reviews
\def\zap{\rf@jnl{ZAp }}                 % Zeitschrift fuer Astrophysik
\def\nat{\rf@jnl{Nature }}              % Nature
\def\iaucirc{\rf@jnl{IAU~Circ.}}       % IAU Cirulars
\def\aplett{\rf@jnl{Astrophys.~Lett.}} % Astrophysics Letters
\def\apspr{\rf@jnl{Astrophys.~Space~Phys.~Res.}}
                % Astrophysics Space Physics Research
\def\bain{\rf@jnl{Bull.~Astron.~Inst.~Netherlands}} 
                % Bulletin Astronomical Institute of the Netherlands
\def\fcp{\rf@jnl{Fund.~Cosmic~Phys.}}  % Fundamental Cosmic Physics
\def\gca{\rf@jnl{Geochim.~Cosmochim.~Acta}}   % Geochimica Cosmochimica Acta
\def\grl{\rf@jnl{Geophys.~Res.~Lett.}} % Geophysics Research Letters
\def\jcp{\rf@jnl{J.~Chem.~Phys.}}      % Journal of Chemical Physics
\def\jgr{\rf@jnl{J.~Geophys.~Res.}}    % Journal of Geophysics Research
\def\jqsrt{\rf@jnl{J.~Quant.~Spec.~Radiat.~Transf.}}
                % Journal of Quantitiative Spectroscopy and Radiative Trasfer
\def\memsai{\rf@jnl{Mem.~Soc.~Astron.~Italiana}}
                % Mem. Societa Astronomica Italiana
\def\nphysa{\rf@jnl{Nucl.~Phys.~A}}   % Nuclear Physics A
\def\physrep{\rf@jnl{Phys.~Rep.}}   % Physics Reports
\def\physscr{\rf@jnl{Phys.~Scr}}   % Physica Scripta
\def\planss{\rf@jnl{Planet.~Space~Sci.}}   % Planetary Space Science
\def\procspie{\rf@jnl{Proc.~SPIE}}   % Proceedings of the SPIE

\let\astap=\aap
\let\apjlett=\apjl
\let\apjsupp=\apjs
\let\applopt=\ao

\title{Hidden star-formation in the cluster of galaxies Abell 1689
\thanks{Based on observations collected at the European Southern
Observatory, La Silla, Chile (ESO No 61.A-0619) }}

   \author{P.--A. Duc \inst{1,2}  \and B.M. Poggianti \inst{3}
   \and  D. Fadda  \inst{4,2}  \and D. Elbaz   \inst{2,5}
   \and  H. Flores  \inst{6,2}  \and P. Chanial \inst{2}
   \and  A. Franceschini \inst{7} \and A. Moorwood \inst{8}
   \and  C. Cesarsky \inst{8}}

   \offprints{paduc@cea.fr}

   \institute{
CNRS URA 2052
\and
CEA, DSM, DAPNIA, Service d'astrophysique, 91191 Gif--sur--Yvette Cedex, France
\and
Osservatorio Astronomico di Padova, vicolo dell'Osservatorio 5, 35122 Padova, Italy
\and
Instituto de Astrofisica de Canarias, Via Lactea s/n, E-38200 La Laguna - Tenerife, Spain
\and
Department of Astronomy \& Astrophysics, University of California, Santa Cruz, CA 95064,USA
\and
DAEC/LUL, Observatoire de  Paris--Meudon, 5 place Jules Janssen, 92195 Meudon, France
\and
Dipartimento di Astronomia, Universit\`a di Padova, Vicolo dell'Osservatorio, 5, I35122 Padova, Italy
\and
European Southern Observatory, Karl-Schwarzchild-Strasse, 2 D-85748 Garching bei M\"unchen, Germany
}
   \date{Accepted for publication in A\&A}

   \titlerunning{Hidden star formation in Abell 1689}

   \authorrunning{P.A.\ Duc et al.}

\abstract{At a redshift of 0.18, Abell 1689  is so far the most distant  
cluster of galaxies  for which
 substantial mid--infrared (MIR) data have been published.
 Its mapping with the ISOCAM camera onboard the ISO satellite allowed the
detection of 30 cluster members at 6.75~\micron\ (LW2 filter) and 16
cluster members at 15~\micron\ (LW3 filter) within a clustercentric
radius of 0.5~Mpc \citep[][ Paper~I]{Fadda00b}.  We present here the
follow--up optical photometric and spectroscopic observations which
were used to study the individual properties of the galaxies members
of A1689. We confirm the high fraction of blue galaxies initially
reported in this rich cluster by \citet{Butcher84}, that was
challenged by some subsequent studies.  
 We discuss the spectral and morphological properties
of all cluster members in our spectroscopic sample, and of the
MIR--detected galaxies in particular. 
Sources with a low [15~\micron] / [6.75~\micron] flux ratio  typically consist of
 luminous passive early--type galaxies while those with a high MIR color index are mainly luminous,
blue, emission--line, morphologically disturbed spirals, i.e. the
star--forming galaxies usually associated with the 'Butcher--Oemler'
 effect. On the other hand, at least 30\% of the 15~\micron\ sources
have optical counterparts showing no evidence of  current
star--formation activity, while their 15 \micron\ emission is most
likely due to obscured star formation.  We argue that the LW3 luminosity
measured in the cluster members is a reliable tracer of the
 total infrared luminosity which in A1689 galaxies peaks at 
$\Lir = 6.2 \x 10^{10}~\Lo$.
 We derive from \Lir\ a star--formation rate free
of dust extinction, SFR(IR), which we compare  with that determined in
the optical from the flux of the \OIIt\ emission line, SFR(opt).
The highest total star formation rates (11~\usfr) and dust extinction  are measured in
 those galaxies exhibiting in their optical spectrum a signature of a dusty
 starburst. 
In contrast, none of the galaxies with post-starburst
optical spectra has been detected by ISOCAM down to a 15 \micron\ flux
limit corresponding to 1.4~\usfr.  
We find a median SFR(IR) of the
LW3--detected galaxies of 2~\usfr, that is ten times higher than the
median SFR(opt) of the \OIIt-- detected galaxies. 
 The ratio SFR(IR)/SFR(opt) is in fact very high, ranging
between 10 and 100 for LW3--detected galaxies with \OIIt\ emission. 
 We conclude that a major part, at least 90\%, 
of the star formation activity taking place in Abell 1689 is hidden. Whether the
 high extinction measured
in the star-forming cluster members results from the cluster
environment itself or reflects a comparable extinction in the coeval
field is still unclear.
}
\maketitle

%-----------------
\section{Introduction}
%-----------------
The history of star formation (SF) as a function of redshift has recently received extensive
 attention since the reference work  by \citet{Madau96}. Combining optical observations
on field galaxies in the nearby ($z=0$), intermediate redshift ($z<1$) and distant ($1<z<4$) 
universe, they  highlighted the strong rise of the star formation comoving density
up to $z=1$ and its  possible decline at higher redshift. Further multiwavelength studies
 have added points on the curve and addressed various biases and uncertainties in the
slopes of the SF rise and decline. In contrast, the global star formation history in clusters is
  practically unknown \citep[See though][]{Kodama01}. A number of impediments make its
study particularly difficult. First of all, data for distant clusters ($z>1$) are still 
missing. Furthermore, variations in cluster properties, such as the X--ray luminosity, 
 the level of sub-clustering and,  generally, the dynamical status of the cluster, 
 introduce a large dispersion in any trend. A ``global'' history -- even just
limited to rich clusters -- can only be obtained using large
samples. Also, the star formation history of clusters is largely related to the star
formation history in the coeval field via the infall rate of field galaxies which is
itself redshift-dependent as well. Any uncertainty in the latter has a strong impact on the
 former. Finally, problems with dust obscuration are critical when computing the
field star formation rate (SFR) density, and they could be at least as annoying in clusters.

The long--term effect of the cluster environment is undoubtedly to quench star--formation
 in its member galaxies \citep[e.g.][]{Abraham96,Poggianti99,Balogh99}. Indeed, nearby rich clusters
 contain more passive, non-starforming
galaxies than more distant ones. This is one aspect of the so--called
Butcher--Oemler effect which is the observed increase in the number of
 blue, presumably star--forming galaxies in clusters, 
as a function of redshift \citep[][ hereafter BO84]{Butcher84}.
 Furthermore, at intermediate redshifts 
the star formation rate per cluster galaxy appears to be lower
than in similar types of galaxies in the surrounding 
field \citep[e.g.][]{Dressler99,Balogh00a}. Unfortunately, 
 estimates of the SFR rely on measurements of optical line fluxes, often \OIIt, obtained with
aperture limited slits. Some \HII\ regions might have been missed and hence the SFR
 underestimated. Data from narrow--band \Ha\ images  have recently
become available for a few $z=0.2-0.3$ clusters \citep{Balogh00b}.
This kind of data has the advantage of
covering the whole galaxy, but it is less sensitive to low levels of star formation
activity. Moreover, the SFR derived in the optical might suffer from strong dust extinction.
 A constant dust absorption of 1 mag at \Ha, typical
of local spirals \citep{Kennicutt92}, is usually assumed to correct
the optically-derived SFR. However, dust obscuration varies dramatically from
one galaxy to another, and it is likely to depend also on the
environmental conditions, as will be discussed in this paper.
Direct optical
signatures of dust enshrouded starbursts have already been found in some cluster
galaxies that have a spectrum classified as, according to the authors, 'e(a)' 
\citep{Dressler99}, 'a+em' \citep{Balogh00b} or 'S+A' 
\citep[in the field, ][]{Flores99b}. All these
spectra show a moderate \OII\ line in emission and strong early Balmer absorption lines. Such 
features are well reproduced by spectrophotometric 
models with a selective dust extinction which
affects differently young and older stars \citep{Poggianti01}.
Though this spectral combination
is able to reveal the occurrence of {\it some} level of hidden star formation,
the models have shown 
that it is impossible to quantify the total SFR on the  
basis of optical observations alone.

Furthermore, several spectroscopic surveys of distant clusters 
\citep[e.g.][]{Couch87,Fabricant91,Barger98,Fisher98,Dressler99} 
have disclosed a significant population of k+a galaxies whose
spectrum is characterized
by the absence of emission lines and the presence of strong Balmer lines
 in absorption. Classically, such objects are considered as post--starburst
or post-starforming
galaxies  \citep[e.g.][]{Couch87,Poggianti99}. Could they however be extreme e(a) galaxies
in which star formation is totally obscured at optical wavelengths, as
suggested by \citet{Smail99}  from  radio-continuum observations? 
In addition, the post--starburst
interpretation of the strongest k+a's requires that a starburst episode occurred in these
galaxies, possibly  when they fell in the cluster. Optical surveys have yet failed
to pinpoint the putative starbursting progenitors, perhaps for
time scale reasons \citep{Couch01}, and ``e(a)'' galaxies have been suggested
to be the most likely progenitors of the post-starburst galaxies
\citep{Poggianti99}.

Undoubtedly deep mid--infrared surveys can address some of the previous questions. 
Fluxes at 10--15~\micron, along with radio centimetric fluxes, provide the best
 estimates of the dust obscured star formation rates \citep[e.g.][]{Elbaz01}
when far--infrared data are not available.
The latest generation of infrared cameras, such as ISOCAM \citep{Cesarsky96} on board of ISO, have
provided for the first time images in the mid--infrared regime with a field of view,
 spatial resolution and sensitivity \citep{Genzel00} particularly well suited
 to the study of distant clusters. 

We made use of the unique capabilities of ISOCAM to
 map at 6.75~\micron\ (LW2 filter,bandwidth:5.0--8.3\AA) and 15~\micron\
 (LW3 filter,bandwidth:11.6--18.0\AA) a sample of ten
 clusters in the redshift range 0.2--0.9 \citep{Fadda00b}. 
The mid--infrared data of the most nearby cluster of our sample, Abell 1689, were
 presented in \citet[][ hereafter Paper~I]{Fadda00}. Towards this rich cluster, situated at a
redshift of 0.18 \footnote{We have taken 
\Hoa\ and \qo=0.5. Adopting this cosmology, the distance of Abell 1689 is 767 Mpc,
which is the value used throughout the paper. With a more typical cosmology, such as
$\Omega_{m}=0.3, \Omega_{\lambda}=0.7$, and \Hob, the distance would become 877 Mpc.
At the adopted distance of Abell 1689, 1 arcmin corresponds to 159 kpc.}, we detected     
 45 infrared sources (41 LW2 sources and 25 LW3 sources). Among them, 6 (13\%) sources
were identified as stars and 8 (18\%) turn out to be foreground or background objects.
A1689 was found to contain a population of galaxies with high
15~\micron\ fluxes and high MIR to optical flux ratios, while such
population is absent towards the central region of the Virgo and Coma clusters.
This was interpreted as revealing an 'infrared Butcher-Oemler effect'.
Moreover, the comparison of optical and MIR counts in the field and 
towards A1689 seemed to suggest a possible excess in the number 
of MIR emitters in the cluster as compared to the number expected
given the different spiral fraction in the two environments and
the fact that most of the LW3 emitters are spiral galaxies.
Such an excess would indicate that SF is triggered in some galaxies
by the cluster environment.
The 15~\micron\ luminosities and MIR-to-optical flux ratios, instead, 
were not found to be significantly different in the cluster 
and in the coeval field.

In the present paper, we focus on the properties of the 
galaxies members of A1689 that were 
detected with ISOCAM, comparing the optical and the IR view of
the Butcher-Oemler effect. A major goal of this work is to establish
the amount of SF hidden by dust in the galaxies of A1689, and
whether a population of   starburst galaxies 
is present in the cluster but would be unrecognized on the basis
of the optical data. 

Our study relies on an extensive spectroscopic survey
of the cluster that has been carried  out with the ESO New Technology Telescope (NTT)
 and on a collection of new and previous photometric data. The 
literature on Abell 1689 is indeed very rich. The cluster was included in the
original paper highlighting what was later known as the Butcher--Oemler effect
(BO84). Several authors have obtained additional photometric
data and discussed whether the rather large blue fraction found in A1689 by
BO84 was real or not \citep[e.g. ][]{Gudehus91,Pickles91,Molinari96,Margoniner00}. 
The lensing properties of the cluster, studied through deep optical images,
 were used to determine its mass profile \citep{Tyson95,Dye01}. The latter was also
estimated from X--ray maps \citep{Miralda-Escude95}.  Evidence
for dynamical substructures was found by \citet{Teague90} based on a set of
spectroscopic data.

The paper is organized as follows: in Section~2  we present our spectroscopic
and photometric observations and compare them with data from the literature.
In Section~3, we describe the properties of the optically-selected
(Sec.~3.2) and MIR-selected cluster members (Sec.~3.3). 
Finally, in 
Section~4, we discuss the star formation activity and the dust extinction of the
galaxies in the cluster and provide some clues on environmental effects that
affected their evolution.

%-----------------
\section{Observations and data-reduction}
%-----------------

\subsection{MIR data}
The observations and data reduction of the ISOCAM mid-infrared data have been described
in detail in Paper~I. We recall here the basic characteristics of the  survey. The total
field of view  was 6' $\times$ 6' (corresponding
to 0.95 $\times$ 0.95 Mpc) with a pixel size of 6". The spatial resolution and
astrometry were good enough to allow an unambiguous optical identification of all the
 mid--infrared sources but the ones near the very crowded cluster center. The 90\% completeness
 limits were 0.2 and 0.4 mJy at 6.75~\micron\ and 15~\micron, respectively.
The sensitivity limits of the LW2 and LW3 surveys were 0.15~mJy and 0.3~mJy, respectively.

\subsection{Imaging data}
B\footnote{The EMMI red channel blue--like Bb filter which was
used has 
a transmission curve slightly different from the standard Johnson's filter.}, 
V and R images of the cluster
 have been obtained during the spectroscopic run which was carried out in May 1998 at the
New-Technology Telescope at La Silla observatory. The seeing was poor  (about 1.5'').
Landolt fields of photometric standard stars (Landolt 1992) were observed for flux
 calibration. Due to the atmospheric conditions, the accuracy of the absolute photometric
 zeropoints is no better than 0.05 mag.
The astrometry of each frame was computed  using several tens of reference stars from the
 USNO A1.0 catalog \citep{Monet96} queried via the ESO SKYCAT browser. The images were
 corrected for distortions during this process. Sources were extracted from the reduced
 BVR images using SExtractor \citep{Bertin96}. The photometric catalog is complete
to B=23~mag, V=22.7~mag and R=22.7~mag. The corrected
 isophotal ('mag\_best') BVR magnitudes from SExtractor are listed in
 Tables~\ref{tab:idMOS}--\ref{tab:idback}.

\begin{table*}[h]
\caption{Spectroscopically-confirmed cluster members in the MOS sample}
\begin{tiny}
\begin{tabular}{lccccccccc}
\hline \tabsp 
ID  & ID$^*$ & RA & DEC & z & B & V & R & morph \\ 
(MOS) & (ISO) & J2000 & J2000 & (MOS) & mag & mag & mag &  \\ 
\tabsp \hline \tabsp 
1 &  -- &  13:11:24.00 &  -1:19:07.6 &  0.1829 &  21.53 &  19.98 &  19.24 &  -- \\
2 &  -- &  13:11:24.25 &  -1:21:14.8 &  0.1766 &  21.41 &  20.82 &  20.27 &  Scd \\
3 &  -- &  13:11:24.27 &  -1:18:38.1 &  0.1803 &  21.87 &  20.38 &  19.65 &  Sb \\
4 &  -- &  13:11:24.40 &  -1:21:11.5 &  0.1870 &  20.72 &  19.11 &  18.42 &  S0 \\
5 &  -- &  13:11:24.41 &  -1:22:15.8 &  0.1785 &  20.48 &  18.81 &  18.08 &  -- \\
6 &  -- &  13:11:25.07 &  -1:19:31.7 &  0.1797 &  21.44 &  19.79 &  19.07 &  S0 \\
7 &  \bf{3} &  13:11:25.31 &  -1:20:37.6 &  0.1924 &  19.75 &  18.13 &  17.38 &  S0 \\
8 &  .. &  13:11:25.64 &  -1:17:24.8 &  0.1814 &  20.91 &  19.31 &  18.59 &  E \\
9 &  -- &  13:11:25.94 &  -1:19:36.0 &  0.1983 &  21.81 &  20.48 &  19.77 &  Sab \\
10 &  -- &  13:11:26.51 &  -1:22:01.4 &  0.1953 &  21.21 &  19.66 &  18.98 &  Sc \\
11 &  .. &  13:11:27.03 &  -1:16:10.5 &  0.1945 &  20.08 &  18.86 &  18.13 &  Sbc \\
12 &  -- &  13:11:27.05 &  -1:18:49.2 &  0.1852 &  22.36 &  20.88 &  20.09 &  Sa \\
13 &  -- &  13:11:27.06 &  -1:21:43.4 &  0.1955 &  22.52 &  21.28 &  20.68 &  S0 \\
14 &  \bf{4} &  13:11:27.07 &  -1:20:58.9 &  0.2153 &  20.81 &  19.68 &  19.10 &  Scd \\
15 &  -- &  13:11:27.37 &  -1:22:48.2 &  0.1832 &  21.13 &  19.43 &  18.73 &  -- \\
16 &  \bf{6} &  13:11:27.69 &  -1:21:06.7 &  0.2165 &  20.59 &  19.50 &  19.06 &  Sd \\
17 &  -- &  13:11:27.80 &  -1:21:13.4 &  0.1977 &  21.55 &  19.95 &  19.19 &  -- \\
18 &  -- &  13:11:27.86 &  -1:21:37.2 &  0.1752 &  21.13 &  19.66 &  18.94 &  S0 \\
19 &  -- &  13:11:27.98 &  -1:18:44.5 &  0.1890 &  21.26 &  19.67 &  18.89 &  S0 \\
20 &  -- &  13:11:28.15 &  -1:18:12.7 &  0.1902 &  22.16 &  20.44 &  19.85 &  S0 \\
21 &  8 &  13:11:28.16 &  -1:20:44.0 &  0.1712 &  20.34 &  18.75 &  18.04 &  S0 \\
22 &  -- &  13:11:28.29 &  -1:18:45.5 &  0.1999 &  21.04 &  19.60 &  18.92 &  -- \\
23 &  \bf{10} &  13:11:28.30 &  -1:19:58.9 &  0.1780 &  20.29 &  18.95 &  18.34 &  Sab \\
24 &  -- &  13:11:28.42 &  -1:22:42.1 &  0.1842 &  20.48 &  18.82 &  17.96 &  -- \\
25 &  -- &  13:11:28.56 &  -1:20:27.1 &  0.1909 &  21.34 &  19.77 &  19.08 &  S0 \\
26 &  -- &  13:11:28.69 &  -1:19:03.3 &  0.1859 &  21.91 &  20.21 &  19.44 &  -- \\
27 &  -- &  13:11:29.26 &  -1:19:17.3 &  0.1932 &  21.30 &  19.57 &  18.89 &  S0 \\
28 &  12 &  13:11:29.27 &  -1:17:50.4 &  0.1726 &  20.34 &  19.24 &  18.79 &  Sab \\
29 &  \bf{14} &  13:11:29.44 &  -1:20:28.4 &  0.1842 &  18.42 &  16.87 &  16.02 &  Scd \\
30 &  .. &  13:11:29.71 &  -1:17:22.1 &  0.1996 &  20.87 &  19.20 &  18.43 &  S0 \\
31 &  -- &  13:11:29.74 &  -1:17:43.2 &  0.1830 &  22.45 &  21.11 &  20.44 &  Sbc \\
32 &  -- &  13:11:29.83 &  -1:20:15.8 &  0.2016 &  21.82 &  20.11 &  19.27 &  Sb \\
33 &  16 &  13:11:29.94 &  -1:20:40.8 &  0.2002 &  19.83 &  18.72 &  18.08 &  Sp \\
34 &  -- &  13:11:29.95 &  -1:22:07.9 &  0.1985 &  20.48 &  18.96 &  18.21 &  S0 \\
35 &  -- &  13:11:29.96 &  -1:20:17.9 &  0.1801 &  21.62 &  19.96 &  19.28 &  S0/a \\
36 &  .. &  13:11:29.98 &  -1:16:25.6 &  0.1785 &  21.17 &  19.56 &  18.84 &  -- \\
37 &  18 &  13:11:30.01 &  -1:20:43.3 &  0.2038 &  19.53 &  18.10 &  17.34 &  -- \\
38 &  -- &  13:11:30.11 &  -1:22:30.8 &  0.1910 &  20.81 &  19.34 &  18.67 &  S0 \\
39 &  20 &  13:11:30.23 &  -1:20:29.8 &  0.1745 &  19.53 &  17.87 &  17.17 &  E \\
40 &  22 &  13:11:30.53 &  -1:20:44.3 &  0.1918 &  20.40 &  18.78 &  17.91 &  E \\
41 &  -- &  13:11:31.03 &  -1:21:28.3 &  0.1878 &  20.28 &  18.50 &  17.87 &  E \\
42 &  24 &  13:11:31.09 &  -1:21:25.7 &  0.1972 &  20.93 &  19.42 &  18.32 &  E \\
43 &  \bf{26} &  13:11:31.36 &  -1:19:33.4 &  0.1878 &  19.54 &  17.81 &  17.03 &  E \\
44 &  -- &  13:11:31.47 &  -1:19:25.5 &  0.1745 &  21.27 &  19.74 &  19.05 &  Sbc \\
45 &  .. &  13:11:31.54 &  -1:17:28.4 &  0.1913 &  20.93 &  19.26 &  18.55 &  S0 \\
46 &  \bf{30} &  13:11:32.07 &  -1:19:47.5 &  0.1801 &  20.37 &  18.43 &  17.72 &  Sa \\
47 &  -- &  13:11:32.17 &  -1:22:11.3 &  0.1855 &  20.57 &  18.90 &  17.93 &  -- \\
48 &  -- &  13:11:32.43 &  -1:22:18.6 &  0.1794 &  21.35 &  20.46 &  19.98 &  ? \\
49 &  \bf{31} &  13:11:32.60 &  -1:18:42.1 &  0.1757 &  19.44 &  18.17 &  17.65 &  Scd \\
50 &  -- &  13:11:32.61 &  -1:18:37.9 &  0.1823 &  -- &  23.96 &  20.84 &  S0 \\
51 &  \bf{32} &  13:11:32.62 &  -1:19:59.3 &  0.2022 &  19.13 &  17.39 &  16.63 &  -- \\
52 &  33 &  13:11:32.67 &  -1:19:32.4 &  0.2009 &  19.55 &  17.66 &  16.83 &  S0 \\
53 &  .. &  13:11:33.08 &  -1:17:02.4 &  0.1891 &  19.94 &  18.29 &  17.58 &  -- \\
54 &  \bf{35} &  13:11:34.02 &  -1:21:02.5 &  0.1813 &  19.73 &  18.49 &  17.94 &  Sa \\
55 &  \bf{37} &  13:11:34.47 &  -1:18:11.7 &  0.1989 &  19.66 &  18.34 &  17.73 &  Sc \\
56 &  .. &  13:11:34.65 &  -1:17:43.7 &  0.1926 &  20.22 &  18.54 &  17.77 &  E \\
57 &  -- &  13:11:34.72 &  -1:20:59.8 &  0.1895 &  20.90 &  19.19 &  18.49 &  E \\
58 &  .. &  13:11:35.10 &  -1:23:18.3 &  0.1925 &  21.42 &  19.90 &  19.37 &  -- \\
59 &  40 &  13:11:35.31 &  -1:21:33.8 &  0.1870 &  20.39 &  18.69 &  17.95 &  -- \\
60 &  .. &  13:11:35.47 &  -1:17:42.9 &  0.1758 &  20.99 &  19.29 &  18.58 &  Sa \\
61 &  \bf{41} &  13:11:35.55 &  -1:20:13.0 &  0.2000 &  19.86 &  18.76 &  18.26 &  Sc \\
62 &  -- &  13:11:36.52 &  -1:18:47.3 &  0.1769 &  22.18 &  20.56 &  19.85 &  E \\
63 &  .. &  13:11:36.58 &  -1:22:54.6 &  0.1758 &  20.15 &  19.08 &  18.45 &  -- \\
64 &  -- &  13:11:37.01 &  -1:22:32.4 &  0.1878 &  21.73 &  20.31 &  19.59 &  -- \\
65 &  .. &  13:11:37.12 &  -1:17:07.7 &  0.1851 &  20.88 &  19.21 &  18.42 &  S0 \\
66 &  -- &  13:11:37.80 &  -1:19:21.2 &  0.1839 &  19.92 &  18.32 &  17.61 &  E \\
67 &  .. &  13:11:37.88 &  -1:22:38.0 &  0.1887 &  21.19 &  19.69 &  18.94 &  -- \\
68 &  .. &  13:11:37.91 &  -1:18:09.0 &  0.1804 &  19.91 &  18.17 &  17.45 &  E \\
69 &  43 &  13:11:38.23 &  -1:21:05.7 &  0.1964 &  20.17 &  18.82 &  18.21 &  Scd \\
70 &  .. &  13:11:39.31 &  -1:16:49.8 &  0.1845 &  20.13 &  18.48 &  17.86 &  E \\
71 &  .. &  13:11:39.56 &  -1:17:50.1 &  0.1977 &  20.86 &  19.28 &  18.51 &  S0 \\
72 &  \bf{45} &  13:11:40.09 &  -1:19:52.3 &  0.1880 &  19.87 &  19.13 &  18.70 &  Sp(late) \\
73 &  .. &  13:11:40.27 &  -1:18:01.1 &  0.1807 &  20.46 &  18.90 &  18.16 &  -- \\
74 &  .. &  13:11:43.38 &  -1:19:20.4 &  0.1837 &  19.83 &  18.01 &  17.29 &  -- \\
\tabsp \hline 
\multicolumn{9}{l}{\parbox{11cm}{$^{*}$ Galaxies detected at 15~\micron\ are
highlighted. Those outside the ISOCAM field of view are indicated with
``..''. Undetected ones in both LW2 and LW3 filters are indicated with ``--''.}} \\
\end{tabular}
\end{tiny}
\label{tab:idMOS}
\end{table*}

\begin{table*}
\caption{Spectroscopically-confirmed cluster members from Teague et al. (1990)}
\begin{tabular}{lcccccccc}
\hline \tabsp 
RA & DEC & ID & z & B & V & R & morph \\ 
J2000 & J2000 & (ISO) &  & mag & mag & mag &  \\ 
\tabsp \hline \tabsp 
13:11:18.37 &  -1:18:40.3 &  -- &  0.2095 &  18.92 &  17.69 &  17.05 &  S0 \\
13:11:20.74 &  -1:20:02.9 &  -- &  0.1835 &  20.47 &  18.83 &  18.07 &  Sa \\
13:11:21.53 &  -1:19:44.2 &  -- &  0.2171 &  20.76 &  19.13 &  18.41 &  -- \\
13:11:26.85 &  -1:19:37.5 &  -- &  0.1754 &  20.34 &  19.02 &  18.38 &  E \\
13:11:27.08 &  -1:22:09.5 &  -- &  0.1840 &  21.17 &  19.41 &  18.63 &  SB0 \\
13:11:27.89 &  -1:23:09.1 &  -- &  0.1842 &  20.48 &  18.77 &  18.02 &  -- \\
13:11:28.96 &  -1:21:17.3 &  -- &  0.1947 &  20.53 &  18.86 &  18.16 &  E \\
13:11:29.04 &  -1:21:37.7 &  -- &  0.1858 &  21.21 &  19.57 &  18.81 &  S0 \\
13:11:29.08 &  -1:21:55.8 &  -- &  0.1908 &  21.02 &  19.35 &  18.57 &  S0 \\
13:11:29.35 &  -1:18:35.4 &  -- &  0.1751 &  20.81 &  19.16 &  18.44 &  -- \\
13:11:30.00 &  -1:20:43.0 &  -- &  0.1987 &  -- &  -- &  -- &  -- \\
13:11:30.20 &  -1:20:28.0 &  -- &  0.1750 &  -- &  -- &  -- &  -- \\
13:11:30.50 &  -1:20:46.0 &  -- &  0.1987 &  -- &  -- &  -- &  -- \\
13:11:30.90 &  -1:20:31.0 &  -- &  0.1739 &  -- &  -- &  -- &  -- \\
13:11:31.00 &  -1:21:28.0 &  -- &  0.1865 &  -- &  -- &  -- &  -- \\
13:11:31.04 &  -1:20:53.1 &  -- &  0.1885 &  21.32 &  19.55 &  18.80 &  E \\
13:11:31.30 &  -1:19:33.0 &  -- &  0.1864 &  -- &  -- &  -- &  -- \\
13:11:32.06 &  -1:21:38.7 &  29 &  0.1770 &  20.57 &  18.98 &  18.33 &  S0 \\
13:11:32.57 &  -1:23:52.1 &  -- &  0.1865 &  20.30 &  18.62 &  17.87 &  -- \\
13:11:33.74 &  -1:18:44.7 &  -- &  0.1849 &  20.79 &  19.08 &  18.37 &  E \\
13:11:35.34 &  -1:20:43.4 &  -- &  0.1835 &  21.21 &  19.54 &  18.80 &  E \\
13:11:35.98 &  -1:23:41.4 &  -- &  0.1861 &  20.29 &  18.54 &  17.76 &  -- \\
13:11:39.43 &  -1:19:07.2 &  -- &  0.1813 &  20.39 &  19.02 &  18.40 &  Sa \\
\tabsp \hline 
\end{tabular}
\label{tab:idNED}
\end{table*}

\begin{table*}
\caption{ISOCAM cluster members with photometric redshift}
\begin{tabular}{ccccccccc}
\hline \tabsp 
ID & RA & DEC & z & B & V & R & morph \\ 
(ISO) & J2000 & J2000 & (phot) & mag & mag & mag &  \\ 
\tabsp \hline \tabsp 
   7 &  13:11:27.78 &  -1:20:08.3 &  0.193 &  20.31 &  18.50 &  17.78 &  E \\
  11 &  13:11:28.40 &  -1:20:25.7 &  0.202 &  21.13 &  19.19 &  18.53 &  E \\
  13 &  13:11:29.36 &  -1:20:43.9 &  0.180 &  20.98 &  19.43 &  18.60 &  Sb \\
\bf{17} &  13:11:29.99 &  -1:20:29.1 &  0.191 &  20.42 &  18.67 &  17.77 &  E \\
  19 &  13:11:30.18 &  -1:20:52.3 &  0.191 &  20.44 &  18.59 &  17.69 &  E \\
\bf{21} &  13:11:30.34 &  -1:20:46.0 &  0.191 &  20.92 &  19.28 &  18.19 &  E \\
\bf{39} &  13:11:35.21 &  -1:18:54.9 &  0.190 &  22.61 &  21.08 &  20.42 &  Sbc \\
\tabsp \hline 
\end{tabular}
\label{tab:idphot}
\end{table*}

\begin{table*}
\caption{Foreground and background galaxies in the MOS sample} 
\begin{tabular}{lcccccccc}
\hline \tabsp 
RA & DEC & ID & z & B & V & R & Morph \\ 
J2000 & J2000 & (ISO) & (MOS) & mag & mag & mag &  \\ 
\tabsp \hline \tabsp 
%13:11:23.95 &  -1:21:46.3 &  -- &  0.1330 &  -- &  21.55 &  21.00 &  Sa \\
13:11:23.95 &  -1:21:46.3 &  -- &  0.1330 &  -- &  21.55 &  21.00 &  ? \\
%13:11:23.95 &  -1:21:46.3 &  -- &  0.1330 &  -- &  21.55 &  21.00 &  Sp(late) \\
13:11:24.58 &  -1:20:04.1 &  -- &  0.4816 &  -- &  22.65 &  21.51 &  Sd \\
13:11:27.19 &  -1:20:10.5 &  \bf{5} &  0.0862$^a$ &  19.41 &  18.24 &  17.60 &  Sab \\
13:11:27.81 &  -1:18:53.3 &  -- &  0.3840 &  -- &  -- &  22.70 &  Sc \\
13:11:28.25 &  -1:18:28.1 &  -- &  0.7220 &  22.94 &  21.67 &  20.91 &  Sd \\
13:11:28.31 &  -1:18:32.7 &  -- &  0.0130 &  21.90 &  21.34 &  20.88 &  E \\
13:11:28.73 &  -1:21:43.9 &  -- &  0.7900 &  21.05 &  20.97 &  20.34 &  Irr \\
13:11:29.67 &  -1:17:47.4 &  \bf{15} &  0.3972 &  20.99 &  19.72 &  18.97 &  Scd \\
13:11:30.73 &  -1:21:39.2 &  \bf{23} &  0.6919 &  21.89 &  21.46 &  20.60 &  Irr \\
13:11:33.00 &  -1:21:25.1 &  -- &  0.1430 &  23.16 &  22.06 &  21.43 &  S0/a \\
13:11:33.69 &  -1:19:39.5 &  -- &  0.3100 &  22.85 &  21.16 &  20.26 &  -- \\
13:11:34.07 &  -1:22:35.8 &  -- &  0.4340 &  -- &  23.01 &  21.86 &  -- \\
13:11:35.18 &  -1:20:30.9 &  -- &  0.5867 &  -- &  22.59 &  21.33 &  Sd \\
13:11:35.95 &  -1:22:29.0 &  -- &  0.2420 &  21.71 &  20.59 &  20.12 &  -- \\
13:11:36.40 &  -1:22:06.1 &  -- &  0.9443 &  -- &  22.23 &  21.59 &  -- \\
13:11:37.37 &  -1:18:37.0 &  -- &  0.0825$^b$ &  19.30 &  18.56 &  18.10 &  Sc \\
13:11:38.76 &  -1:19:08.2 &  -- &  0.3695 &  22.05 &  20.61 &  19.97 &  Sd \\
13:11:41.83 &  -1:19:48.1 &  -- &  0.1030 &  21.19 &  19.94 &  19.35 &  -- \\
\tabsp \hline \tabsp
\multicolumn{8}{l}{\parbox{11cm}{Notes:$^a$ [TCG90] 217; the redshift of 0.2153
as given in \citet{Teague90}, is presumably wrong. $^b$ [TCG90] 006; the redshift
 of 0.1826, as given in \citet{Teague90}, is presumably wrong. }} \\
\end{tabular}
\label{tab:idback}
\end{table*}

Our photometric database also includes the Gunn g, r and i magnitudes measured by
 \citet{Molinari96} with the ESO 3.6m telescope and independently by \citet{Margoniner00}
with  the CTIO 0.9m telescope,  the B and I magnitudes measured by \citet{Dye01}
 at the Calar--Alto 3.5m telescope  and the
near-infrared K'--band magnitudes measured by \citet{DePropris99} at the CTIO 1.5m.
All these different catalogs were  cross-identified. Unfortunately because  these
 data have various origins and were extracted with different methods,
 they cannot be easily compared. In particular they turn out  to be of little use
to derive a reliable spectral energy distribution.
In the following, we will thus mostly rely on our own BVR photometric data.

Finally, archive HST/WCPC2 images (PI:Tyson) covering the whole MOS field of view have
 been processed by the MORPHS group and were kindly given to us. They consist of
 recombined F555W and F814W images.
The HST astrometry was lost during the combination process and it was re-computed  using the astrometric
solution of our ground-base images as reference.

\subsection{Spectroscopic data}
 The spectroscopic observations were carried out  with EMMI installed
 on the NTT. Using five  masks punched before the observations, we  obtained
 spectra of 111 different objects towards the  inner $\rm Mpc^2$ 
(5' $\times$ 7') of Abell 1689. Each slitlet was 1.3'' wide (3.4~kpc at the distance of A1689)
and at least 8'' long. The slit length was actually adjusted to cover the whole length
 of the target object keeping some blank sky at each side. 
 Note that, due to geometrical constraints, the slit orientation was not  
necessarily  along the main axis. 
The targets were selected from a deep V-band image of the
cluster which was available in the NTT archive. We chose preferentially the optical counterparts
 to the MIR sources detected by ISOCAM (Paper I) and completed the gaps between the slitlets
 with bright cluster candidates. The completeness of the MOS sample is analyzed in Appendix~A.
 It is shown  that the initial selection criterion results only in a slight bias towards
blue galaxies.
Whenever possible, objects in the same magnitude range were arranged in each mask.
The source lists of cluster members and fore-/background galaxies
are given in Table~\ref{tab:idMOS} and \ref{tab:idback}, respectively.
The total integration time was about 2 hours per mask, divided in 4 exposures of 30 minutes each.
The disperser, ESO grism\#3, had a resolution of about 700 at 6000~\AA\ . The wavelength range
 depended on each slitlet position on the mask. We made sure that the redshifted
4000~\AA\ break was present in all spectra. The wavelength of \Ha\ is available only for a few
 galaxies whereas the \OII\ position is reached in most cases.

Data reduction and extraction were performed using a set of IRAF procedures written by PAD.
A normalized dome flat was used as a flat-field. The wavelength calibration, based on 
HeAr lamp spectra, was carried out on the 2D spectra that were initially roughly re-positioned 
to a common wavelength using the info of the slitlet positions available in the headers.
The spectra of several spectrophotometric standard stars were obtained with the same grism
and  5'' wide long-slits.
A few objects were observed through different masks with their corresponding slitlets
put at various locations. We could hence check the relative accuracy of the flux
calibration in the MOS field, which turns out to be better than 20\%.
 
Redshifts were determined from the average value of individual emission and absorption lines
  with a stronger weight given to emission lines. For  spectra with very
low signal to noise, no obvious lines could be identified; instead the wavelength of the 4000~\AA\
 decrement was used. The redshift of thirty galaxies could be compared with 
that measured by \citet{Teague90}. They are identical (within 1\%) for all of them, but two 
\footnote{[TCG90] 006 and [TCG90] 217 have a redshift of resp. 0.0825 and 0.0862 instead of
0.1826 and 0.2153 as reported in \citet{Teague90}. Given the number of spectral features, we believe
our redshifts are the correct ones.}. Our redshifts are listed in Table~\ref{tab:idMOS} and
Table~\ref{tab:idback}.

Line measurements were performed with two techniques: manually, with the gaussian fitting provided
by the 'splot' procedure in IRAF, and semi-automatically with a purposely written program \citep[MORPHS collaboration, ][]{Dressler99}.
 They agree well with each other, although the manual technique underestimates the equivalent widths by
15--20\%. This is due to systematic differences in the adopted level of the continuum. In order to
compare our results with those of the MORPHS group, we decided to base our spectral classification
and analysis on the measurements obtained with the semi-automatic method. 
For galaxies observed through different masks, we retained the spectra with the highest signal
 to noise. 
The spectrophotometric data of our MOS run are listed in Table~\ref{tab:spec}.

\begin{table*}
\caption{Spectrophotometric data of cluster members from the MOS sample}
\begin{tiny}
\begin{tabular}{lccccccl}
\hline \tabsp 
ID & ID & ID & flux(\OII) & -EQW(\OII) & EQW(\Hd) & Class & Comments \\ 
(MOS) & (ISO) & (mask) & \uflux & \AA & \AA & &  \\ 
\tabsp \hline \tabsp 
1 &  -- &  MOS5:17 &  -- &  -- &  -- &  k: & poor  \\
2 &  -- &  MOS2:9 &  4.67\er0.19 &  32.8\er1.7 &  4.8:\er 1.0 &  e(c) & \Hb=-8.2,OIII=-14.0  \\
3 &  -- &  MOS3:20 &  -- &  -- &  -- &  ? &  poor  \\
4 &  -- &  MOS1:12 &  -- &  -- &  3.8:\er 2.1 &  k+a: & poor \\
5 &  -- &  MOS1:6 &  -- &  -- &  -- &  k: &  OIII=-1.5? \\
6 &  -- &  MOS3:17 &  -- &  -- &  -- &  k: & poor  \\
7 & \bf{3}  &  MOS5:11 &  0.60\er0.32 &  3.9\er1.6 &  -- &  k(e) &   \\
8 &  .. &  MOS1:31 &  -- &  -- &  -- &  k: & poor  \\
9 &  -- &  MOS4:17 &  -- &  -- &  3.8\er1.4 &  k+a &   \\
10 &  -- &  MOS1:7 &  -- &  -- &  -- &  k: &   \\
11 &  .. &  MOS1:36 &  2.73\er0.28 &  8.6\er1.0 &  2.3\er0.7 & e(c:) &  \Hb=-2.9,OIII=-9.4,OIII2=-2.7 \\
12 &  -- &  MOS3:19 &  -- &  -- &  -- &  ? & poor  \\
13 &  -- &  MOS2:6 &  -- &  -- &  -- &  ? & poor  \\
14 & \bf{4} &  MOS4:10 &  1.85\er0.27 &  18.8\er3.3 &  -- &  e(c) & \Hb=-3.6  \\
15 &  -- &  MOS5:3 &  -- &  -- &  -- &  k: &   \\
16 & \bf{6} &  MOS5:9 &  2.36\er0.11 &  60.0\er3.0 &  -- &  e(b) &  \Hb=-12.8,OIII=-44.4 \\
17 &  -- &  MOS5:8 &  -- &  -- &  -- &  k &   \\
18 &  -- &  MOS5:7 &  -- &  -- &  -- &  k: &  OIII/sky? \\
19 &  -- &  MOS5:18 &  -- &  -- &  -- &  k &   \\
20 &  -- &  MOS2:22 &  -- &  -- &  -- &  k: & poor  \\
21 &  8 &  MOS5:10 &  -- &  -- &  -- &  k &   \\
22 &  -- &  MOS4:21 &  -- &  -- &  -- &  k: & poor  \\
23 & \bf{10} &  MOS2:15 &  3.05\er0.23 &  10.4\er1.0 &  2.1:\er 0.8 &  e(a)+ & $<\rm H\theta + H\eta + H\zeta>/3 = 5.1$, \Hd\ prob.higher,\Hb=-0.8\\
24 &  -- &  MOS1:4 &  -- &  $<3.3$  &  -- &  k:(e:) &  \Hb=2.0 \\
25 &  -- &  MOS3:13 &  -- &  -- &  -- &  k: &   \\
26 &  -- &  MOS4:19 &  -- &  -- &  -- &  ? &   \\
27 &  -- &  MOS4:18 &  -- &  -- &  3.0:\er 1.3 &  k+a: &   \\
28 &  12 &  MOS1:29 &  -- &  -- &  5.5\er0.7 &  k+a &   \\
29 & \bf{14} &  MOS2:13 &  -- &  -- &  -- &  k &   \\
30 &  .. &  MOS2:25 &  -- &  -- &  2.1\er0.9 &  k &   \\
31 &  -- &  MOS5:21 &  -- &  -- &  -- &  k: & poor  \\
32 &  -- &  MOS4:14 &  -- &  -- &  6.7:\er 2.4 &  k+a: & poor  \\
33 &  16 &  MOS4:11 &  -- &  -- &  -- &  k: & poor  \\
34 &  -- &  MOS5:5 &  -- &  -- &  -- &  k &   \\
35 &  -- &  MOS5:12 &  0.23:\er 0.09 &  7.4:\er 2.3 &  -- &  k(e:) &   \\
36 &  .. &  MOS1:35 &  0.53:\er 0.18 &  2.1:\er 1.1 &  4.2:\er 1.7 &  k+a:(e) &  NII in em? \\
37 &  18 &  MOS4:12 &  -- &  $<5.4$  &  1.8\er0.7 &  k(e::) &   \\
38 &  -- &  MOS4:3 &  -- &  -- &  2.8\er1.1 &  k+a: &   \\
39 &  20 &  MOS1:15 &  -- &  -- &  2.0:\er 0.6 &  k &  \\
40 &  22 &  MOS3:12 &  -- &  -- &  -- &   k: &  poor \\
41 &  -- &  MOS2:7 &  -- &  -- &  -- &  k &   \\
42 &  24 &  MOS1:11 &  -- &  -- &  -- &  k &   \\
43 & \bf{26} &  MOS1:20 &  -- &  -- &  -- &  k &   \\
44 &  -- &  MOS5:16 &  -- &  -- &  -- &  k: &   \\
45 &  .. &  MOS3:22 &  -- &  -- &  -- &  k: &  poor \\
46 & \bf{30} &  MOS4:16 &  -- &  -- &  -- &  k: &   \\
47 &  -- &  MOS2:4 &  -- &  -- &  -- &  k &   \\
48 &  -- &  MOS3:5 &  1.50\er0.17 &  46.2\er7.0 &  -- &  e(b) &  \Hb=-6.7,OIII=-11.4,\Ha=-43.1 \\
49 &  \bf{31} &  MOS1:24 &  2.48\er0.26 &  7.1\er0.9 &  5.5\er0.8 &  e(a) &  $<\rm H\theta + H\eta + H\zeta>/3=5.2$  \\
50 &  -- &  MOS1:25 &  -- &  -- &  -- &  ? &  very poor \\
51 &  \bf{32} &  MOS1:18 &  -- &  -- &  1.1\er0.8 &  k &   \\
52 &  33 &  MOS2:16 &  -- &  -- &  2.0\er0.5 &  k &   \\
53 &  .. &  MOS1:32 &  -- &  -- &  -- &  k &   \\
54 &  \bf{35} &  MOS1:13 &  1.90\er0.17 &  8.1\er0.8 &  3.5\er0.7 &  e(a)+ &$<\rm H\theta + H\eta + H\zeta>/3=5.5$, Ha+NII=-3.1 \\
55 &  \bf{37} &  MOS1:27 &  1.00\er0.30 &  3.9\er1.0 &  2.8\er1.4 &  e(a)+ & $<\rm H\theta + H\eta + H\zeta>/3=5.7$, Ha+NII=-13  \\
56 &  .. &  MOS1:30 &  -- &  -- &  2.3:\er 1.6 &  k(e:) &  \Ha=-2.3? \\
57 &  -- &  MOS3:11 &  -- &  -- &  -- &  k: &  poor \\
58 &  .. &  MOS1:2 &  0.18:\er 0.07 &  2.5:\er 0.9 &  3.4\er1.2 &  k+a(e:) &   \\
59 &  40 &  MOS3:9 &  -- &  -- &  -- &  ? &  very poor \\
60 &  .. &  MOS4:24 &  -- &  -- &  -- &  k+a: &  earl Bals v. str (to be meas) \\
61 &  \bf{41} &  MOS2:14 &  -- &  -- &  3.5\er0.8 &  sey1: & OIII=-5.1,broad Hb em=-3.7,broad Ha+NII*2=-28.4 \\
62 &  -- &  MOS2:19 &  -- &  -- &  -- &  ? &  poor \\
63 &  .. &  MOS3:3 &  .. &  .. &  -- &  e(c) & sky on OII,no \Hd,OIII=-16.8,\Hb=-6.5,\Ha=-9.1  \\
64 &  -- &  MOS1:5 &  -- &  -- &  -- &  k: &  poor \\
65 &  .. &  MOS4:25 &  -- &  -- &  -- &   k &  no \\
66 &  -- &  MOS2:17 &  -- &  -- &  2.5\er1.0 &  k &   \\
67 &  .. &  MOS3:4 &  -- &  -- &  -- & k(e) &  NII2=-4.8 \\
68 &  .. &  MOS3:21 &  -- &  -- &  4.1:\er 2.2 &  k+a: &   \\
69 &  43 &  MOS2:10 &  .. &  .. &  2.9\er1.0 &  e(c) & \Hb=1.4,Ha$>$-4.2,NII*2  \\
70 &  .. &  MOS2:26 &  .. &  .. &  -- &   k &   \\
71 &  .. &  MOS5:20 &  .. &  .. &  -- &   k &  \Hb=1.9 \\
72 &  \bf{45} &  MOS3:16 &  .. &  .. &  -- &  e(b) &  OIII=-11.7,\Hb=-15.4,Ha$>$-42.7,NII2 \\
73 &  .. &  MOS4:23 &  .. &  .. &  -- &   k: &   \\
74 &  .. &  MOS1:21 &  .. &  .. &  -- &   ? &  poor \\
\tabsp \hline \tabsp
\multicolumn{8}{l}{\parbox{\textwidth}{Notes: ``..'' means that the \OIIt\ or \Hd\ line is outside
 the MOS wavelength range. The equivalent width of a number of some other lines are indicated
 in the comments.}} \\
\end{tabular}
\end{tiny}
\label{tab:spec}
\end{table*}

%-----------------
\section{Results}
%-----------------
\label{sec:results}

\subsection{Cluster membership and velocity distribution}

\begin{figure}
\centerline{\psfig{file=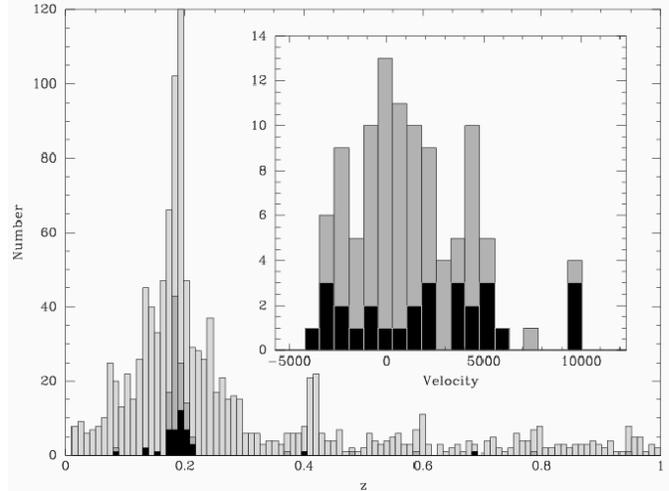,width=\columnwidth}}
\caption{ Redshift histogram towards  Abell 1689 in the redshift range $z=$ 0 -- 1.
Several samples are included: optically selected galaxies with photometric redshifts
 from the catalog by  Dye et al. (2001) (light shaded bars) or spectroscopic redshifts
 from our MOS survey and  from the literature (shaded bars); ISOCAM--selected galaxies 
 with either spectroscopic or photometric redshifts (dark shaded bars).
The inset shows a close up on the redshift range $z=$ 0.17 -- 0.22, that
considered for cluster membership. Only  ISOCAM--selected galaxies with
well determined spectroscopic redshifts are included in the latter plot. 
 The x--axis is scaled in \kms. } 
\label{fig:histz}
\end{figure}

\begin{figure}
\centerline{\psfig{file=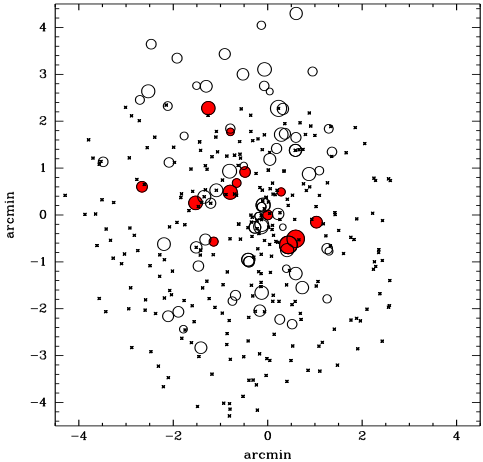,width=\columnwidth,angle=-90}}
\caption{Spatial and velocity distribution of cluster members. Photometrically
 confirmed members are shown with the small stars.  Spectroscopically
confirmed members are shown with circles the size of which is proportional
to  their velocity in the cluster. Finally, the filled (red) circles correspond
to the ISOCAM 15\micron\ sources.}
\label{fig:idvel}
\end{figure}

The histograms shown in Figure~\ref{fig:histz} collect 
all redshift information available for galaxies lying in the 
 5' $\times$ 7' field towards Abell 1689. Spectroscopic redshifts
 mainly come from our MOS survey  and from the
spectroscopic survey by  \citet{Teague90}.
In addition, \citet{Dye01} determined the photometric redshifts of 
 several hundreds of galaxies observed  through 
an optimized set of  narrow--band and broad--band filters.
Not surprisingly, all histograms show a strong concentration of
galaxies in the redshift range 0.17 -- 0.22, presumably members
of Abell 1689. The peak is at $z_{\rm spectro}=0.184$. While the 
 large dispersion in the redshift distribution based on  the
photometric technique results from the uncertainty of the method,
the substantial velocity spread indicated by the spectroscopic
observations for galaxies related to A1689 is real
 (see inset in Fig.~\ref{fig:histz}).
It reflects both the richness of the cluster and its complex
velocity structure. Applying a multiscale analysis,
\citet{Girardi97} distinguished three distinct groups in A1689 that
overlap spatially but are well separated in velocity. 
Hence, what we see in projection towards the central
region of this cluster (see Fig.~\ref{fig:idvel}) is probably composed of
 different   subclumps merging along the line of sight. 

Figure~\ref{fig:histz} indicates that the ISOCAM--detected galaxies 
also show  a well-defined peak in correspondence with the cluster
 baricentric velocity. Actually,  more than 80\% of the
 ISOCAM sources have redshifts between 0.17 -- 0.22.
This prominent excess of mid--infrared emitters related to the cluster
is real and does not result from selection effects in the optical
follow-up (see Paper I). However, the velocity distribution of ISOCAM
 sources related to A1689 is flatter than that of the
 bulk of the cluster population (see inset in  Fig.~\ref{fig:histz}).
 This result is actually expected. We will show later
 that most MIR emitters are associated with spiral and emission--line galaxies.
 Studies of nearby clusters indicate a larger velocity dispersion for this
population than for the early-type galaxies \citep[e.g.][]{Biviano97}.  

The observation of a well defined peak  in the redshift histogram 
 at $z = 0.17--0.22$   both in the optically selected and mid-infrared selected 
samples provide  a redshift--based criteria to assess the cluster membership.
In the following analysis, we consider as cluster members 
{\it at large} all galaxies in the above--mentionned redshift range.
They might be long standing  members, galaxies just falling in the
cluster or belonging to merging sub-clusters. Their excess of infrared activity
as detected by ISOCAM  might trace the broad cluster environment, or
 the cluster itself. A detailed analysis of the precise role of the
environment goes beyond the limits of this study for lack of statistics and
field coverage. Our survey  only extends out to a clustercentric radius
of 0.5 Mpc, therefore inferring a  radial dependence of the 
properties of the cluster members is not possible.

\subsection{Properties of the  optically selected cluster members}
In this section we present the photometric, spectroscopic and morphological
properties of galaxies in our spectroscopic sample,  comparing them
when possible with the properties of lower and higher redshift clusters. 

\subsubsection{The optical subsamples}
The 74 cluster members in our MOS survey are listed in Table~\ref{tab:idMOS}.
In the following analysis we  include another 17 galaxies for which 
redshifts are available in the literature \citep{Teague90}.
These are listed in Table~\ref{tab:idNED}. We also make use of the 167
galaxies  with an R band magnitude brighter than 22,  which have a photometric redshift
 determined by \citet{Dye01} compatible with a cluster membership.
Those detected by ISOCAM are listed in Table~\ref{tab:idphot}.
For reference, we indicate in Table~\ref{tab:idback} the foreground and background 
galaxies found in our MOS survey.

\subsubsection{Blue fraction}
The value of the blue fraction, \fb, in Abell 1689, i.e. the proportion of galaxies bluer than the
 color--magnitude relation, is controversial. 
In their original article, \citet{Butcher84}
 computed a value of $\fb = 0.09 \pm 0.03$.
 However, later on \citet{Gudehus91} estimated a lower value for \fb ($0.05\pm 0.06$)
consistent with the blue fraction measured in local clusters. Recently, \fb\ was revised
again and raised to: $0.093\pm0.019$
 \citep{Margoniner00} and even  $0.191 \pm 0.015$  \citep{Margoniner01}.
Such a discrepancy can be due to 
the many observational biases, the fuzziness of the 
definition of the blue fraction (i.e. the choice of the color index, 
luminosity range,
 k-correction model, location of the color-magnitude diagram etc.)
and especially the background subtraction. 

We  estimated \fb\ using our dataset of confirmed cluster members 
(see details in Appendix~B). Our computed  value   is 1.5--2 times that
 originally derived by \citet{Butcher84} for this cluster and appears to be much higher than in
 the local rich clusters studied by these authors. Large blue fractions have
also been reported in other clusters at z=0.2 \citep[e.g. in Abell 115,][]{Metevier00},
but there is clearly a large spread in \fb\ among rich 
clusters at this redshift \citep{Smail98}.

\subsubsection{Spectral and morphological classification}
We classified all spectra using the set of rules proposed by \citet{Dressler99}, which are
based on the rest-frame equivalent widths of the \OII\ emission line and 
the \Hd\ absorption line \footnote{We have introduced an additional spectral class,
the ``k(e)'' type, which represents spectra similar to the k-type but with
signs of at least one very weak emission line. The definition of the ``e(a)'' class
 was also slightly modified, as indicated and justified in Sec.~3.3.4. Finally, seven
 galaxies for which the \OII\ line was outside the MOS wavelength range were 
classified based on their Hydrogen Balmer lines.}.
No spectral type could be assigned to the spectra of 8 cluster members
due to their low signal to noise.

\begin{figure}
\centerline{\psfig{file=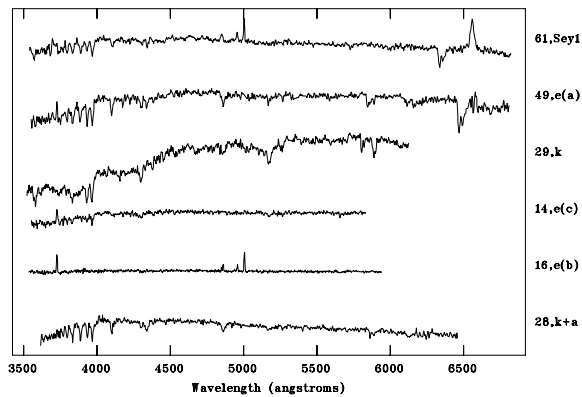,width=\columnwidth}}
\caption{Representative optical spectra of cluster members with different spectral
 types.
All spectra except the bottom one belong to LW3-detected galaxies.
From bottom to top, spectra are ordered by increasing 15~\micron\ flux.
The MOS ID and spectral type
 are indicated to the right.}
\label{fig:spec}
\end{figure}

\begin{figure}
\centerline{\psfig{file=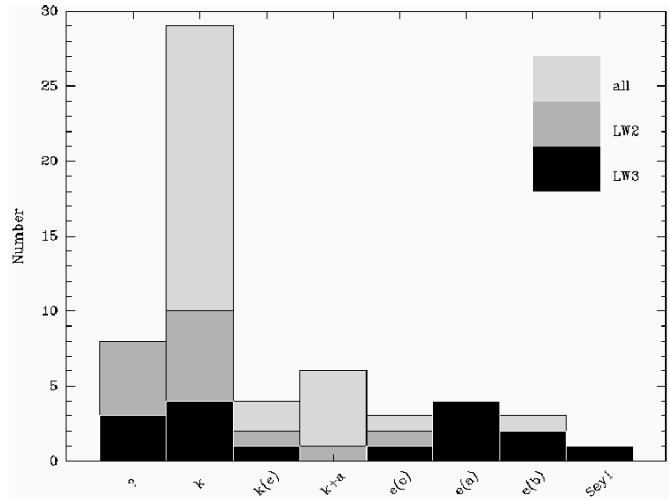,width=\columnwidth}}
\caption{Spectral type histogram of all MOS (light shaded),
6.75~\micron\ detected (shaded,LW2) and 15~\micron\ detected (dark
shaded,LW3) cluster members in the ISOCAM field of view.
Note that almost all LW3 sources have also been detected in
the LW2 filter. The unknown class '?'
either includes  galaxies for which the signal to noise was too low to
determine a redshift or ISOCAM sources missed by our MOS survey.}
\label{fig:histstype}
\end{figure}

Representative spectra for each class are presented in Figure~\ref{fig:spec} and the histogram
 of the spectral types is shown in Figure~\ref{fig:histstype}.
At least 50\% of the spectroscopically confirmed members have spectra of type 'k' -- they show
no emission lines and only weak Balmer absorption lines -- and are hence typical of
 passive early type galaxies.
The 'k+a' type, characterized by the presence of strong Balmer absorption lines typical of 
post--starburst/post-starforming galaxies, accounts for 10 to 15\% of 
our spectroscopic sample. This fraction 
is significantly higher than that typically estimated in nearby clusters 
\citep[$\sim 1$\% according to][]{Dressler87}.
A direct comparison with higher redshift clusters is hindered by the
different completeness as a function of galaxy magnitude of the various spectroscopic
samples. In clusters at $z \sim 0.5$, the k+a fraction was found to be
about 20\%  \citep[MORPHS collaboration, ][]{Dressler99,Poggianti99}. 
Using a sample with a mean redshift 
lower than the MORPHS, the CNOC1 group \citep{Balogh99} estimated a much lower 
proportion of post-starburst galaxies, less than 5\%.
On the other hand, \citet{Abraham96} counted in Abell 2390 a proportion of \Hd--strong
 galaxies as high as 23\%,  in good
agreement with Abell 1689 where the  \Hd--strong objects (k+a's and e(a)'s) amount to
about 20\%. The two clusters are situated at the same redshift.

In total about 20\% of the cluster members in our sample are emission line galaxies 
showing some level of star formation in the optical.
Conversely, most of the galaxies in the cluster classified as 'blue'
  -- those responsible for the photometric Butcher-Oemler effect -- 
 exhibit emission lines  that are indicative of a current star formation.

The spectroscopic sample includes only one Seyfert 1 galaxy, identified by the 
broad \Ha\ emission
line. Usual diagnostics to distinguish Seyfert 2/LINERs from HII regions \citep{Veilleux87} 
could not be applied because the required \Ha\ hydrogen, \NIIt\ nitrogen and \SIIt\ sulfur lines
were  outside the spectral range of most galaxies. Instead, we have used the recent diagnostics of
\citet{Rola97}, based on lines in the blue only. We found that no AGN activity was required
to explain the optical spectra of the emission line galaxies but the
 Seyfert~1. 
 
Finally, we note that all spectral types are evenly distributed throughout 
our field of view.
Star--forming galaxies are present even at low projected clustercentric radii.
The field-of-view  and the 
sub-clustering and related projection effects (see Sec.~3.1)
are likely to be the reasons why
no clear radial trend is actually observed in any of the galactic properties
(optical color, spectral characteristics, morphology).

The morphologies of galaxies towards Abell 1689 were derived and kindly provided to us
by the MORPHS group from HST WFPC2 images, and 
are included in Table~\ref{tab:idMOS}. A description of the  classification 
method may be found in \citet{Smail97}.
 The fraction of E-type galaxies 
does not change if  photometric or spectroscopic members
are considered,  
showing that  our MOS survey is not biased towards/against 
any particular Hubble type of galaxies.
 Within our spectroscopic sample,  the majority (60\%) of 
the members have morphologies
 typical of early type galaxies (ellipticals and lenticulars).
We find a quite large fraction of lenticulars (about 35\%)
and an even larger fraction of spirals (at least 40\%)\footnote{The fractions do not change 
if only galaxies with $M_V<-20$ are included. This magnitude cut is the same
as the one applied in other studies of galaxy morphologies, including
Fasano et al. (2000). The cluster area sampled here  is similar to
the area used in these other works (the central $\rm Mpc^2$).   A comparison
of the morphological fractions in A1689 with those given by these
other authors is therefore justified.}. 
The latter is significantly larger than the spiral fraction in rich clusters at z=0 
\citep[e.g. see Fig.~9 in][]{Fasano00}.

To conclude, for many aspects, the galaxies in Abell 1689 -- despite their low redshift --
seem to have different optical properties than those of galaxies in typical rich clusters in 
the nearby universe. Several of the evolutionary trends associated
 with the Butcher-Oemler effect, i.e. a  change in the photometric, spectroscopic and  morphological
 properties of the cluster members, may already be seen in this $z=0.18$ cluster.

\subsection{Properties of the MIR selected cluster members}
\label{sec:propmir}
Abell 1689 is the first cluster at moderate distance that has been mapped in the
mid-infrared regime with enough sensitivity to compute statistics on the properties of
the MIR emitters.
In the following, we will focus our analysis on the mid-infrared emitters in the cluster
and discuss more in detail the IR side of the Butcher-Oemler effect that has been
reported in Paper I.

\subsubsection{The MIR subsamples}
The ISOCAM cluster sample consists of 30  galaxies detected at 6.75~\micron\ 
with a mean LW2 flux of 0.23 mJy and 16 
 detected at 15~\micron\ with a mean LW3 flux of 0.57 mJy 
\footnote{The galaxy ISO \#28 which, with a photometric redshift of 0.15,
was considered as a possible cluster member in Paper I has not been included here.}.
 Among them, 15  sources have been detected at both  wavelengths. 
 For most MIR sources, the cluster membership has been assessed using spectroscopic
redshifts. Seven MIR sources have been missed in our MOS survey and were considered
as cluster members based on their photometric redshift only. They are listed in
Table~\ref{tab:idphot}.
Five LW3 sources have a low signal to noise  at 15~\micron\ but were included in
 the statistics because the reliability of the detection was  confirmed  by the 
discovery of a strong  6.75~\micron\ counterpart (see Paper I). 
The LW2 sources towards the crowded central regions suffer severe blending
 problems. They were deblended into several sources and their relative LW2 fluxes 
were assigned based on the relative luminosities of the optical counterparts candidates
 (Paper~I).

\subsubsection{Luminosity and color}
\begin{figure}
\centerline{\psfig{file=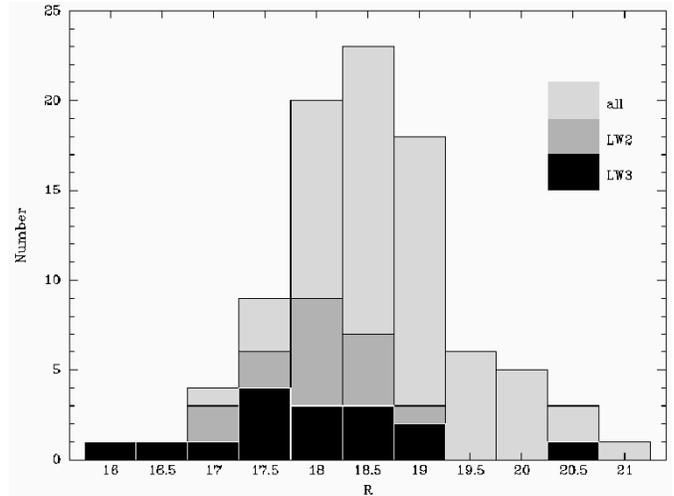,width=\columnwidth}}
\caption{R--band magnitude histogram  of spectroscopically confirmed  (light shaded),
 6.75~\micron\ detected (shaded,LW2) and 15~\micron\ detected (dark
shaded,LW3)    cluster members within the ISOCAM field of view.}
\label{fig:histR}
\end{figure}

\begin{figure}
\centerline{\psfig{file=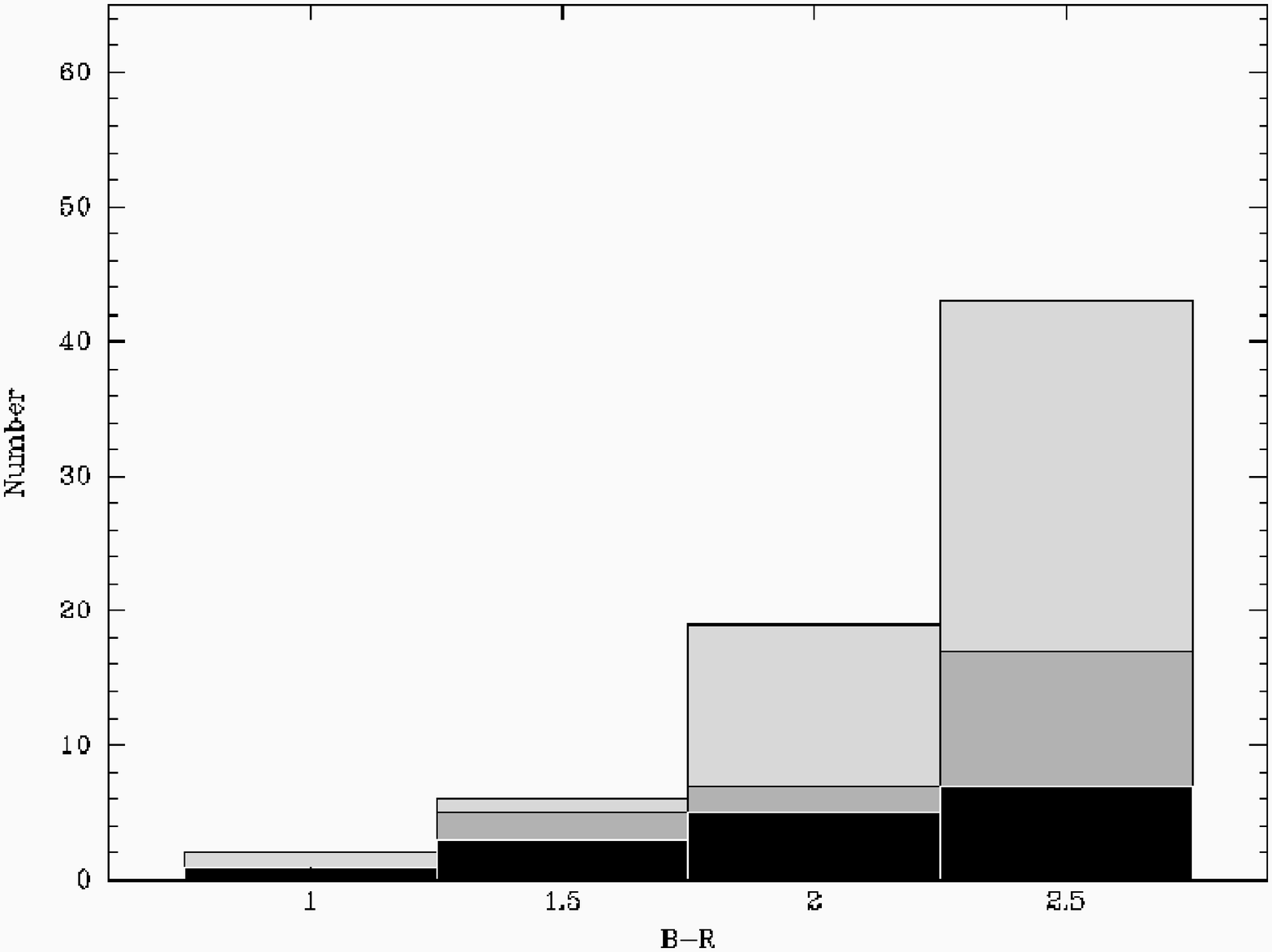,width=\columnwidth}}
\caption{$B-R$ color index histogram of
  spectroscopically confirmed  (light shaded),
 6.75~\micron\ detected (shaded,LW2) and 15~\micron\ detected (dark
shaded,LW3)    cluster members within the ISOCAM field of view.}
\label{fig:histBR}
\end{figure}

\begin{figure}
\centerline{\psfig{file=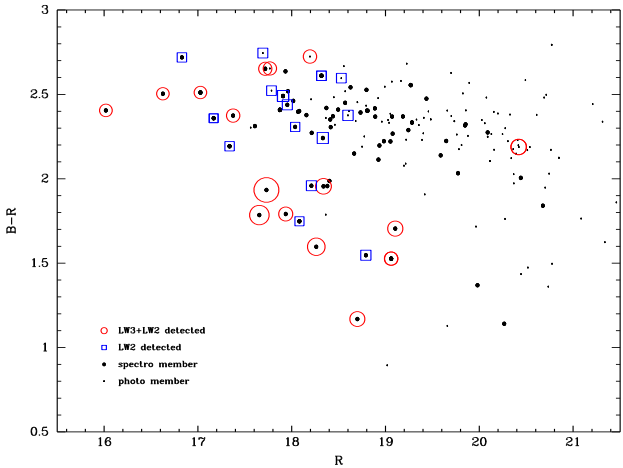,width=\columnwidth,angle=-90}}
\caption{Color--magnitude diagram towards Abell 1689.
 Galaxies whose cluster membership  has been assessed from their photometric
redshift are shown with the small black point. Spectroscopically confirmed members are
shown with the large black points. Galaxies detected at 6.75 (LW2) and 15~\micron\ (LW3)
are indicated by resp. the (blue) squares and the (red) circles.
The size of the latter symbols are proportional to the ISOCAM flux color LW3/LW2.
 Note that almost all LW3 sources were also detected in the LW2 filter.
 For LW2--only detected sources, an upper limit for the LW3 flux was used to
 compute the MIR color.
Only galaxies in the ISOCAM field of view are shown.}
\label{fig:CMRlw}
\end{figure}

The R--band magnitude and B--R color distributions of the cluster MIR emitters are
 shown in Figure~\ref{fig:histR} and Figure~\ref{fig:histBR},
superimposed on the histograms of the optically selected cluster members. Clearly,
 the LW2 sources correspond to the brightest and reddest galaxies in the cluster
while the LW3 sources have a slightly flatter optical luminosity function and
moreover span a large color range. These results may be better visualized in 
the color-magnitude diagram shown in Figure~\ref{fig:CMRlw} where galaxies detected
 at 6.75 and 15~\micron\ are indicated by resp. a square and a circle. The size of 
each symbol is proportional to the LW3/LW2 flux ratio. The majority of the bright
blue galaxies in the cluster 
-- those responsible for the photometric BO effect --
 turn out have a large 15~\micron\ to 6.75~\micron\ flux ratio. In the magnitude range
18.5--19.5 only blue galaxies are detected through the LW3 filter.

\subsubsection{Morphology}
\begin{figure*}[ht]
\centerline{\psfig{file=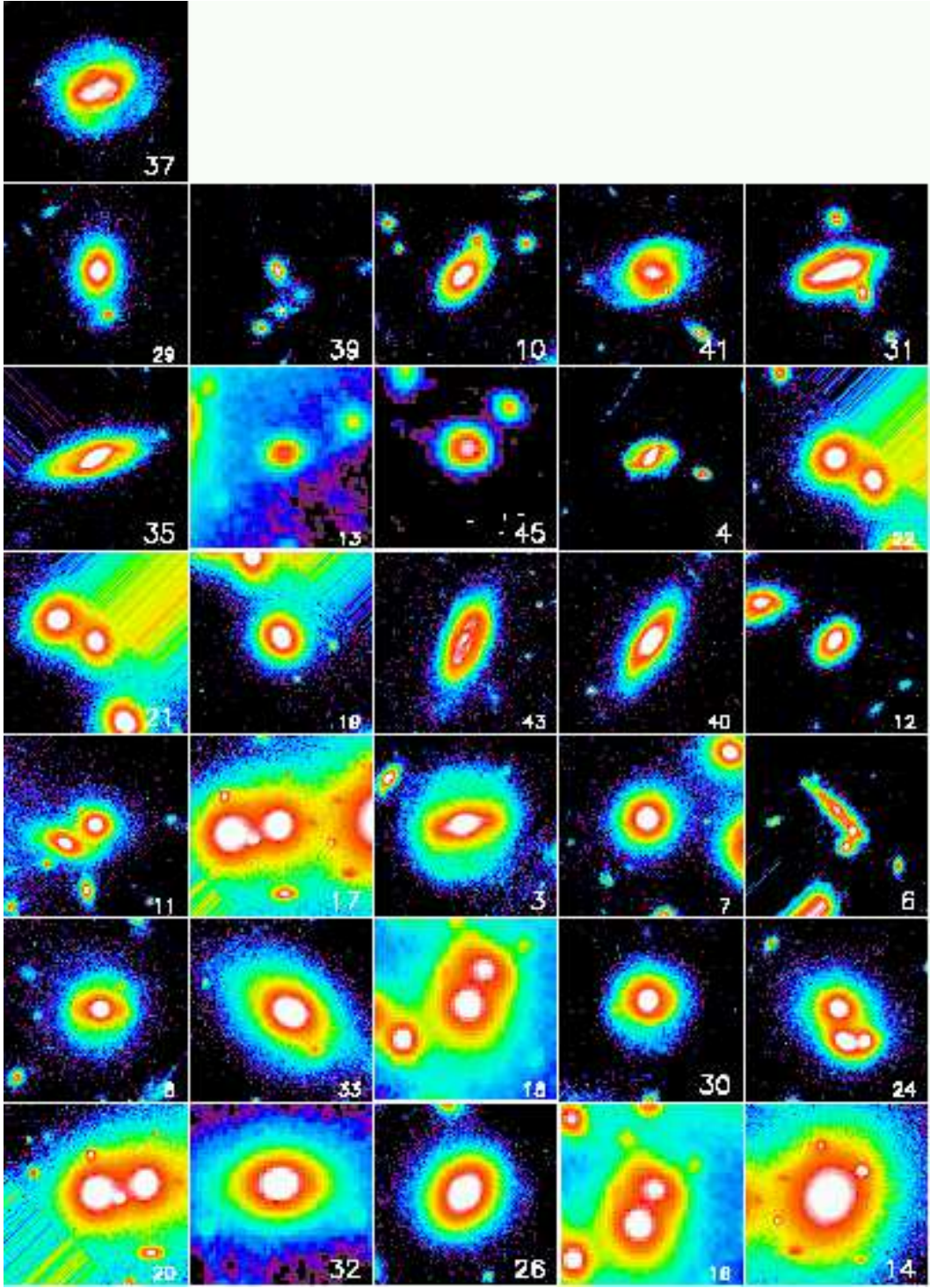,width=14cm}}
\caption{HST/WFPC2 images of ISOCAM detected cluster members. They are ordered
 according to the increasing LW3/LW2 flux ratio (from the lower left to the upper right panels).
The spatial and intensity (logarithmic) scales are the same in all panels. For
ISO\#32 and ISO\#45 which lie outside the HST survey field of view, NTT
 images  are shown instead. Galaxies are labeled with the ISO ID number. Labels with a 
low--size text font indicate objects only detected at 
6.75\micron. For these, an upper limit for the LW3 flux was used to
 compute the MIR color.}
\label{fig:lw3morph}
\end{figure*}

HST morphologies are available for almost all ISOCAM sources. Figure~\ref{fig:lw3morph} displays the optical
 images of the MIR emitters.
These flux--calibrated images are all shown with the same spatial and intensity scale 
and are ordered according to the MIR color LW3/LW2.

First of all, we note a  correlation between the MIR color and the
 size and luminosities of the optical counterparts,  sources with a low
LW3/LW2 ratio being generally bigger in size
and more luminous in the optical.
About half of the LW2 sources have morphologies characteristic of 
early type galaxies (E and S0s). Their LW2 light is expected to be
contaminated by the large integrated
stellar photospheric emission. 
 The galaxies with the highest LW3/LW2 ratio and the majority (75\%)  of the LW3 emitters are spiral
 galaxies. This is  expected 
since most of the emission at rest-frame 12.5 \micron\ arises from star--forming regions
which are prevalent in late-type galaxies. We note however that a few galaxies 
-- 3 galaxies or 19\%, which is already significant -- of LW3 sources have morphologies
 typical of elliptical galaxies, and 1 is an S0. 
Among the galaxies with a high LW3/LW2 ratio,  several of them have clearly a disturbed morphology:
for example, perturbed external disks (ISO source \#37, the strongest MIR emitter in the sample),
 isophote twisting (ISO\#4), weak tidal tail (ISO\#41) or interacting
galaxies (ISO\#6 and, possibly, \#39). 
Morphologically peculiar galaxies are known to be
more frequent in ISO selected samples than in optical
samples \citep{Flores99b}. It is also well known that the most extreme 
far--infrared emitters -- the ULIRGs --
are all mergers \citep{Sanders96b}. However,
as we will show in Section~\ref{sec:sfr}, none of the MIR sources in Abell 1689 have far
IR luminosities as high as ULIRGs.

The galaxy at the core of the X-ray cluster emission, ISO\#14, which is detected at
 15\micron, has an intriguing morphology: its isophotes are ellipticals but
its central surface brightness is magnitudes below that of elliptical or cD galaxies.

\subsubsection{Spectral class}
The dominant
population among  sources with a low LW3/LW2 flux ratio are passive, k--type, galaxies while 50--70\%  of
the LW3--bright sources are emission--line galaxies (see Fig.~\ref{fig:histstype}).
The only AGN  in our MOS sample (ISO\#41) is detected in both ISOCAM bands, with
a LW3/LW2 flux of 4 which is not unusual for an AGN--dominated galaxy \citep{Laurent00}.

One spectral class of particular interest for MIR selected samples is the so--called
'e(a)' class: objects having moderate [OII] emission line and a strong \Hd\ absorption line.
 Objects with such a spectral signature in the optical have been interpreted by
\citet{Poggianti99} as dusty starbursts. Observationally, e(a) is the typical spectral
class of  very luminous infrared galaxies  \citep{Poggianti00} which are notoriously
 powered by dust enshrouded starbursts. 
In this survey, only one galaxy has 
an e(a) spectrum according to the strict $\rm H\delta$ threshold
as defined by \citet{Dressler99}. However,
other three spectra display similar characteristics, with weak to moderate
\OIIt\ emission and unusually strong early Balmer lines in absorption. 
The average rest-frame $<\rm H\theta + H\eta + H\zeta>/3$ of these galaxies is always 
higher than 5 \AA $\,$ (see Table~\ref{tab:spec}) and 
is much stronger than in the spectra of all the other spectral classes, except the k+a's.
In these galaxies the $\rm H\delta$ line might be slightly more filled
by emission than in the ``proper'' e(a) galaxies, but given the strong
$<\rm H\theta + H\eta + H\zeta>/3$ their spectrum is likely to be
the result of star formation and dust properties similar to those in e(a)'s, 
and we decided to include them in the e(a) class hereafter.
e(a) spectra represent about 25\% of our LW3 sample,
 whereas {\it all} e(a)'s in the spectroscopic 
sample are LW3 sources. Once again, this seems to confirm the existence of a connection
between dusty star--forming galaxies and optical spectra with strong early
Balmer lines and weak emission.
The star formation rates and dust extinction
of the e(a) galaxies will be presented in
Section~4.1.3, where it will be shown 
that the SF activity in e(a)'s
is the highest of all spectral classes.

 Surprisingly, 
five  LW3 sources do not show any sign of current or recent star formation
activity in their optical spectra. 
These galaxies have the lowest LW3 fluxes in the sub-sample and
the origin of their LW3 emission will be discussed in Section~4.1.2. 
In total, less than 15\% of  the passive (k-type)  galaxies 
and no post-starburst (k+a) galaxies are detected at 15~\micron .

\subsubsection{The IR side of the BO effect}
The properties of the galaxies from our MIR samples can roughly be 
summarized as  follows: 
 sources with a low LW3/LW2 flux ratio typically consist of luminous, red, passive early
 type galaxies, while those with a 'red' MIR color are mainly luminous, blue, emission line,
 disturbed spiral galaxies  -- precisely
 the galaxies responsible for the optical Butcher-Oemler effects. 
 Moreover, a small but significant fraction of LW3 sources
show no sign of current nor recent SF in their optical spectra.

\section{Discussion}

In the previous section, it has been shown that the majority of LW3 sources
have spectra with emission-lines indicative of ongoing star
formation. Thus, \it qualitatively \rm the optical
SF estimators generally agree with the MIR data in identifying starforming
galaxies, in about 70\% of the cases. 
In this section, we will compare the SFR derived from the \OIIt\ line
and from the MIR flux in a \it quantitative \rm manner, 
with the purpose of evaluating how much of 
the SF activity remains undetected from the spectra.

\subsection{Total star formation rates in Abell 1689}
\label{sec:sfr}

\subsubsection{Optical star formation rate}
Balmer Hydrogen lines are commonly used  to derive star--formation rates in the
optical. However for most of the cluster members, the \Ha\ line was outside our
spectral domain. The \Hb\ emission line is not reliable because it is polluted by
 the absorption line from A stars. Our spectral resolution is too low to properly
subtract this contribution. Measurements in the UV, that are heavily affected
by dust extinction as well, are not available. 
The only tracer at our disposal in the optical is the \OII\ 
emission line. Several studies have emphasized  the many problems of that line
  \citep[e.g.][]{Jansen01,Charlot01}. Being in the blue, it is 
very sensitive to extinction. Moreover, its strength is not simply proportional to the
 number of ionizing photons; it depends very much on the metallicity/excitation of
 the ISM. As a result, the uncertainties in the SFR derived from \OIIt\ may be as
high as a factor of 10.  Nevertheless, this line is used very often as a SF indicator,
given the lack of alternatives in many studies.  

\OII\ could be measured in the spectra of 9 galaxies which are members of A1689 lying within
 the field of view of the ISOCAM survey.
We have used the conversion of \citet{Kennicutt98a}:
\[SFR([OII]) = 1.4 \x 10^{-41} L([OII]/\ul)~\usfr \]
  which is valid for solar 
abundance and a Salpeter IMF. The SFR derived in A1689 galaxies
ranges between 0.05~\usfr\ and 0.5~\usfr\ with an average and median among
\OIIt--detected galaxies of 0.2~\usfr\ per galaxy (see Table~\ref{tab:sfr}).
These values are lower limits, given that no extinction correction  was
 applied. Moreover,  a fraction of the \OIIt\ flux from the galaxies might have
been missed by the slitlet (see Sec.~4.2.1).
After correcting  for  a canonical extinction in the optical of 1 mag 
at \Ha\ \citep{Kennicutt92}, the average SFR per star--forming galaxy rises to 0.5~\usfr. 
The highest optical SFR measured in our sample, 1.15~\usfr, appears  rather modest with
 respect to that found in other clusters at similar redshifts
 \citep[e.g. AC 114 at z=0.32][]{Couch01} and to the average SFR measured in
the coeval field  \citep[e.g.][]{Tresse98} with slit spectroscopy.

\begin{table*}
\caption{Star formation rates of star--forming galaxies in A1689}
\begin{tabular}{lcccccc}
\hline \tabsp 
ID$^a$ & ID$^b$ & $F_{15}/F_{K'}$ $^c$ & $L_{IR}$$^d$ & SFR(IR)$^e$ & SFR(\OIIt)$^f$ \\ 
(MOS) & (ISO) &  --  & $10^{10}~\Lo$ & \usfr & \usfr \\ 
\tabsp \hline \tabsp 
2 &   -- &    ..     &  $<$0.81 &  $<$1.4 &  0.46\er0.02 \\
7 &    3 & 0.5\er0.2 &  1.32 &  2.2\er0.9 &  0.06\er0.03 \\
14 &   4 & 1.4\er1.0 &  0.93 &  1.6\er1.2 &  0.18\er0.03 \\
16 &   6 &    ..     & 0.81: &  1.4:      &  0.23\er0.01 \\
23 &  10 & 1.6\er0.5 &  1.79 &  3.0\er1.0 &  0.30\er0.02 \\
29 &  14 & 0.1\er0.1 &  1.35 &  2.3\er0.9 &  $<$0.05 \\
.. &  17 & 0.3\er0.1 &  1.26 &  2.1\er1.0 &  .. \\
.. &  21 & 0.5\er0.4 &  0.96 &  1.6\er1.2 &  .. \\
43 &  26 & 0.2:      & 0.81: & 1.4: &  $<$0.05 \\
46 &  30 & 0.2:      & 0.81: & 1.4: &  $<$0.05 \\
48 &  -- & ..        &  $<$0.81 &  $<$1.4 &  0.15\er0.02 \\
49 &  31 & 0.9\er0.3 &  1.79 &  3.0\er1.0 &  0.24\er0.03 \\
51 &  32 & 0.1:      &  0.81: &  1.4: &  $<$0.05 \\
54 &  35 & 0.8:      &  1.17: &  2.0: &  0.19\er0.02 \\
55 &  37 & 2.9\er0.4 &  6.19 &  10.5\er1.7 &  0.10\er0.03 \\
.. &  39 & 0.9\er0.5 &  1.08 &  1.8\er1.1 &  .. \\
61 &  41 & 2.7\er0.6 &  2.38 &  4.0\er0.9 &  $<$0.05 \\
69 &  43 & $<$ 0.7   &  $<$0.81 &  $<$1.4 &  .. \\
72 &  45 & 6.2\er1.1 &  3.78 &  6.4\er1.2 &  .. \\
\tabsp \hline \tabsp
\multicolumn{6}{l}{\parbox{10cm}{Notes: this table includes 'active' cluster members in the
ISOCAM field of view that either
have emission lines in their optical spectra or that are detected at 15~\micron.
$^a$ '..' indicates that no MOS spectra is available. $^b$ '--' indicates that the
galaxy has not been detected by ISOCAM. $^c$ 15~\micron\ to 2.2~\micron\ flux ratio.
K' magnitudes were taken from \citet{DePropris99}.
 $^d$ Total IR luminosity estimated from the 15~\micron\ luminosity.
We used the following formulae:
$ \log(\nu L_{\nu}[12\micron] = 0.494 + 0.955 * \log(4 \pi D^{2} \nu F_{\nu}[15\micron]) $
(conversion to the equivalent rest-frame 'IRAS' luminosity at 12\micron)
$ L_{ir} = 0.89* (\nu L_{\nu}[12\micron])^{1.094} \Lo $
\citep[calibration of ][]{Chary01}.
The luminosities indicated with ':' have large errors and may be
upper limits (see Paper I).
 $^e$ Infrared star formation rate derived from $L_{IR}$. For galaxies
 with $F_{15}/F_{K'}$ below 0.1, the
photospheric emission becomes greater than 50\%, and the estimated SFR is 
overestimated by a factor of 2. Above 0.5, the correction is negligible.
 The AGN activity in MOS ID\#61   might contribute to the MIR luminosity,
and hence its derived SFR might be overestimated.
  $^f$ Star formation rate estimated
from the \OIIt\ luminosity, not corrected for any optical extinction
or slit aperture correction.
}} 
\end{tabular}
\label{tab:sfr}
\end{table*}

\subsubsection{Mid infrared emission and star formation activity}
\label{sec:irsfr}
The MIR emission, despite its complexity, is arguably a reliable tracer of the star
 formation activity.
The many different components at the origin of the emission in the ISOCAM bands 
have been discussed in detail in many papers \citep[e.g.][and references  therein]{Genzel00}. In short,
 the main contributors are (a) unidentified infrared bands (UIBs) from photodissociation
 regions (PDR), (b) continuum emission by warm small grains heated by young stars or an AGN
(c) continuum emission from the photosphere of evolved stars, and (d) emission lines from
 the ionized interstellar gas. The exact contribution of each component
is difficult to be assessed using  broad--band photometric data only. When no spectrum
 is available,
the LW3/LW2  color index provides a rough indication of the relative importance
 in the MIR regime of the starburst activity versus more quiescent spiral-like
 star formation, AGN activity and stellar emission, in the sense that a higher
LW3/LW2 generally indicates a more prevalent starburst activity
 \citep[see the empirical models of ][]{Laurent00}.
 Of course, this single diagnostic is at some level 
degenerate and some a priori assumption on the origin of
 the MIR emission should be introduced. 

In our sample, nuclear activity does not contribute
 much.  Only one AGN, of Seyfert 1 type, is present
 in our optical spectroscopic database (see Sec.~3.2.3).
Moreover, it is unlikely that Abell 1689 harbors  obscured AGNs similar to
those discovered in  the most  extreme ULIRGs \citep{Sanders88a}. The contribution
 of the nuclear activity seems to become prominent only in galaxies with a total
 infrared luminosity that  exceeds $10^{12.3}~\Lo$ \citep{Veilleux99,Tran01} while,
as we will see in the following, all the galaxies in our sample have IR
luminosities more than an order of magnitude lower.

\begin{figure}
\centerline{\psfig{file=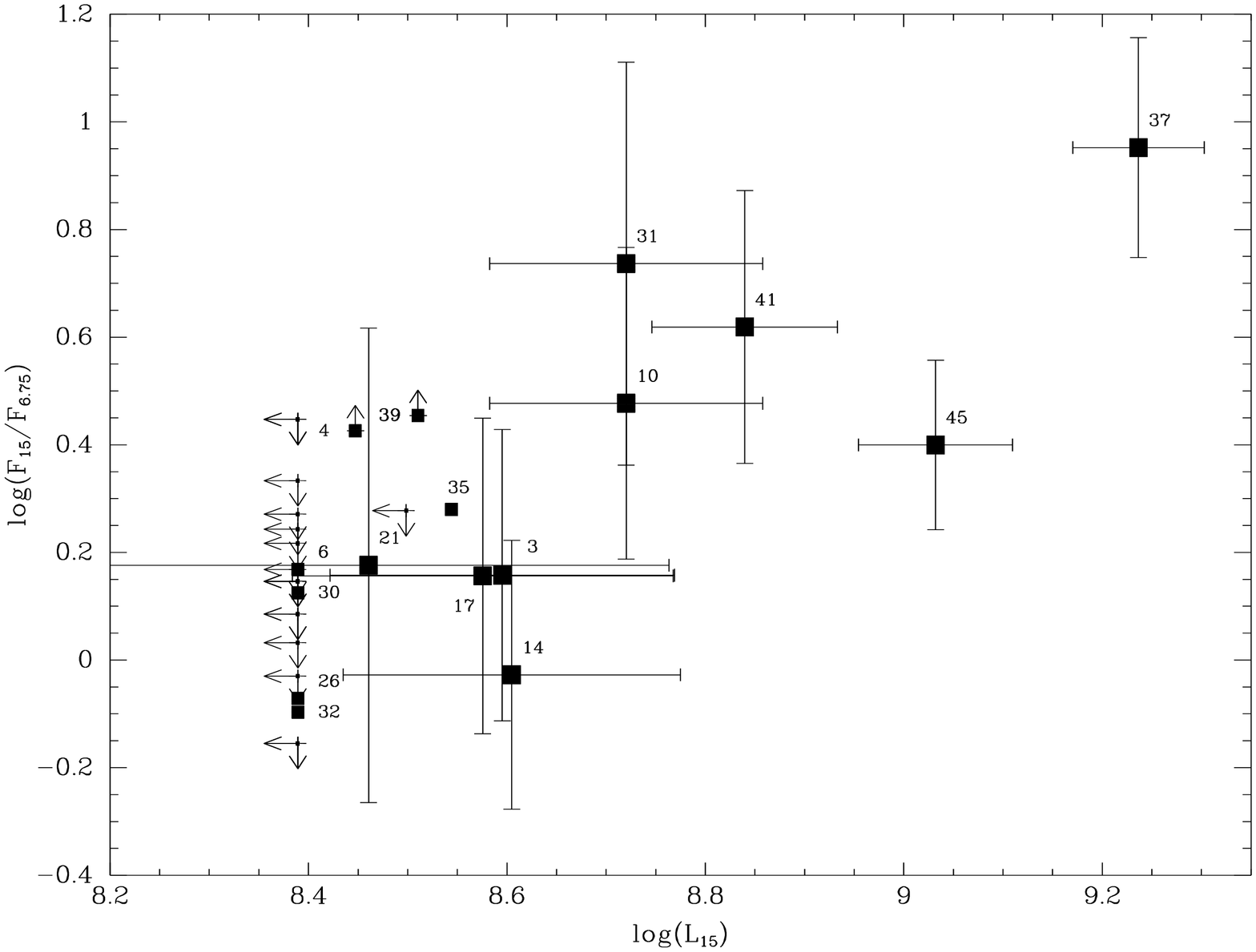,width=\columnwidth,angle=-90}}
\centerline{\psfig{file=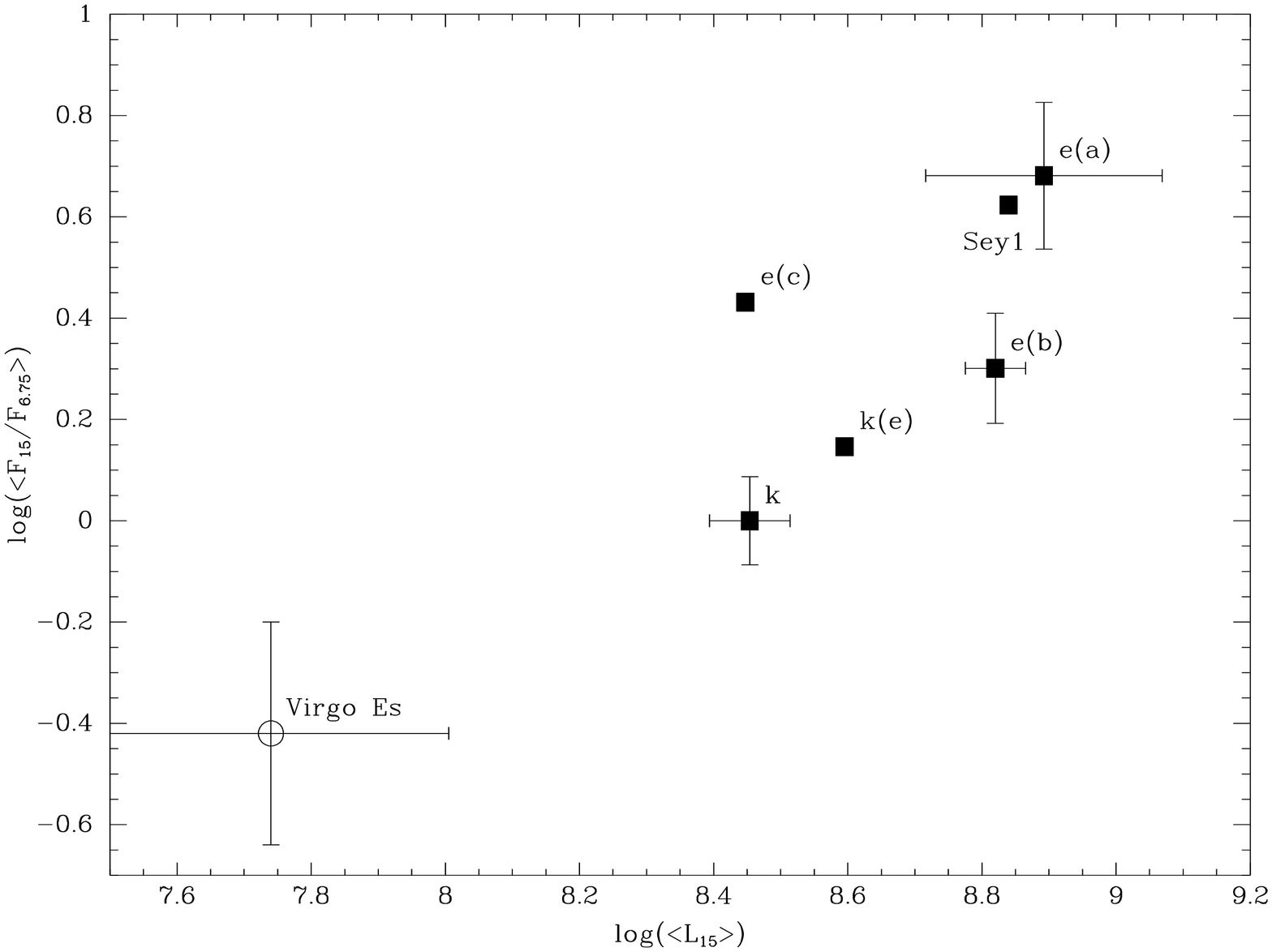,width=\columnwidth,angle=-90}}
\caption{Mid--infrared LW3/LW2 color index versus 15~\micron\ luminosity 
(in \Lo) for ISOCAM  cluster members. Top: individual values. The ISO ID number is
 indicated for galaxies detected at both 6.75~\micron\ and  15~\micron. Bottom: averaged
values per spectral class.  The errors are the deviation to the mean. Only galaxies
detected in both ISOCAM bands were used. The open circle 
corresponds to the average values of early--type galaxies in the Virgo
cluster \citep{Boselli98}. The 15~\micron\ luminosities were computed using the
formula $L_{15} = 4 \pi D^{2} F_{15}\delta_{\nu}$ where $\delta_{\nu}$ is the
LW3 filter width.} 
\label{fig:lw3vlw2}
\end{figure}

\begin{figure}
\centerline{\psfig{file=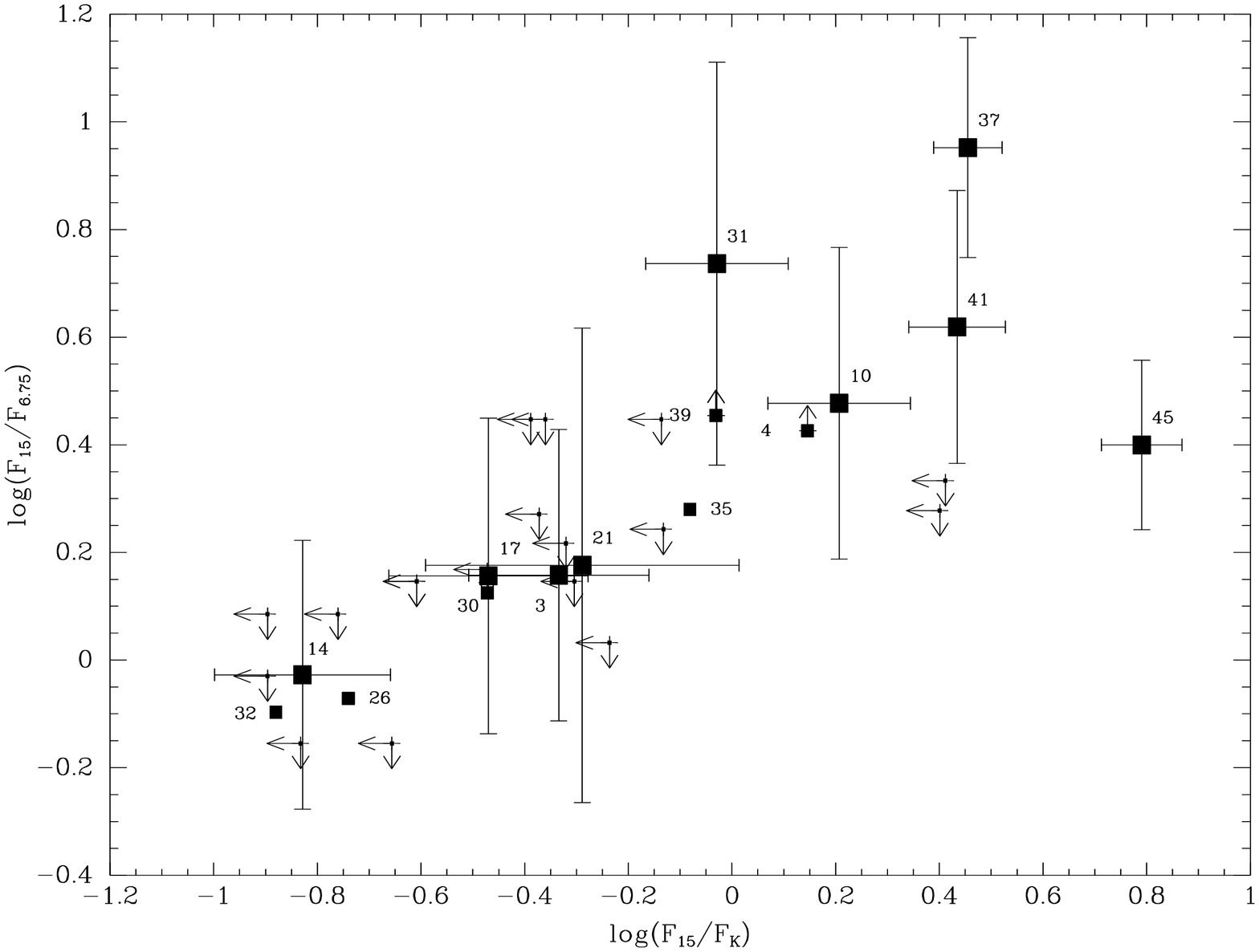,width=\columnwidth,angle=-90}}
\centerline{\psfig{file=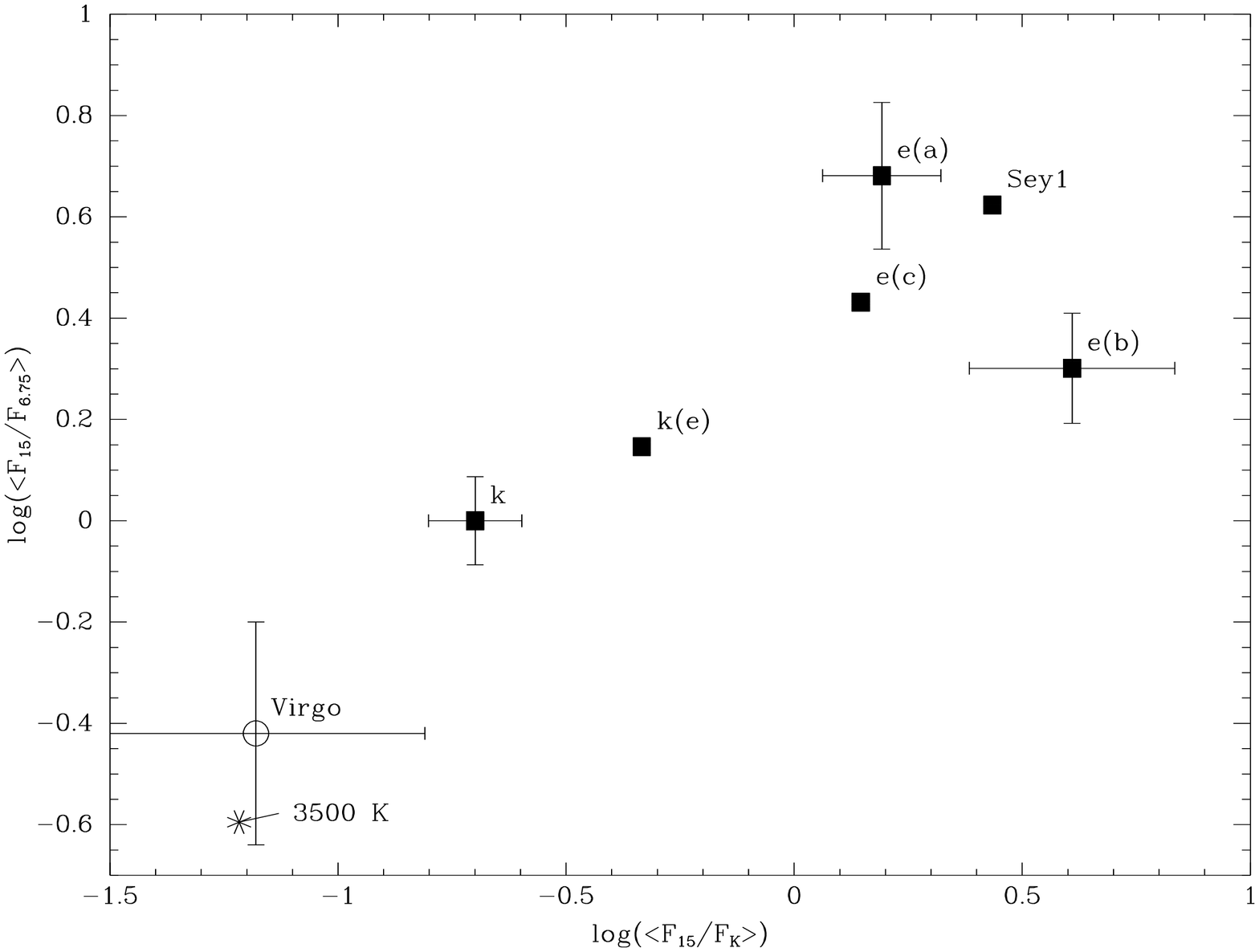,width=\columnwidth,angle=-90}}
\caption{Mid--infrared LW3/LW2 color index versus LW3/K' flux ratio for ISOCAM  
cluster members. Top: individual values. The ISO ID number is indicated for galaxies detected
 at both 6.75~\micron\ and  15~\micron. Bottom: averaged
values per spectral class.  The errors are the deviation to the mean. Only galaxies
detected in both ISOCAM bands were used. 
The star indicates the position 
of a 3500K blackbody which approximates the SED of ellipticals. The associated
arrow shows the k--correction to apply at $z=0.18$. The open circle 
corresponds to the average values of early--type galaxies in the Virgo
cluster \citep{Boselli98}.}
\label{fig:lw23vlw3k}
\end{figure}

 A correlation between  the LW3/LW2 flux ratio and the LW3 luminosity, shown in Figure~\ref{fig:lw3vlw2},
  is observed for the A1689 cluster members, especially when one takes into account the upper
limits.
This trend 
can be {\it qualitatively} understood considering that the LW3 luminosity is  an indicator of the
 star formation activity \footnote{In contrast, the LW2 flux observed at the redshift of Abell 1689 
(rest--frame 5.7~\micron\ ) is mostly due to stellar emission and is
hence a worse tracer of the star formation activity.
Indeed, the LW3/LW2 ratio and the LW2 flux are not correlated.}. 
Passive galaxies have a mid-infrared spectrum in which the Rayleigh--Jeans
 tail of the dominant old stellar component prevails; for a pure stellar
contribution approximated by a $T = 3500~$K black body redshifted to $z=0.2$,  
the MIR color index is as low as log LW3/LW2 = -0.6. 
The MIR spectrum of galaxies with a low to moderate star formation rate,
 typical of normal spirals, is dominated by the UIBs; their MIR color index is typically 
 log LW3/LW2 = 0 to 0.5  at $z=0.2$ \citep[see Fig.~11 in][]{Laurent00}. Finally, the MIR spectrum of
 galaxies  with a high SFR, i.e. starbursts,  is dominated for $\lambda > 10~\micron$ by continuum emission from
 hot dust; their LW3/LW2 ratio increases to values as high as log LW3/LW2 = 1. 
The correlation between the MIR color index and the 15~\micron\ flux is kept when the
latter is normalized by the near--infrared K' flux at 2.2~\micron, a tracer of the total stellar luminosity
(see Fig.~\ref{fig:lw23vlw3k}). Pure stellar MIR emitters have log LW3/K' = -1.1
(after k--correction)
while in dusty active galaxies this ratio may reach  log LW3/K' = -0.5.
The  diagrams presented in Figures~\ref{fig:lw3vlw2} and \ref{fig:lw23vlw3k} show that all LW3--detected
 galaxies in A1689 have a 15~\micron\ flux
which is much higher than expected for a    Rayleigh--Jeans emission. The
  contribution from photospheric stellar emission may be up to 50\% for galaxies with the lowest
 LW3/K' flux ratio, 
on average is about 20\%, and less than 2\% for the 15~\micron--bright galaxies. 

Therefore the 15~\micron--detected population   is composed of star--forming cluster members
with  some (5 to 7) galaxies with LW3/LW2 ratios in the range typical for normal spirals,
while the remaining (3 to 5) galaxies have higher LW3/LW2 ratios suggestive
of a current starburst.

\subsubsection{Infrared star formation rate}
The level of star--formation activity may be estimated in a more {\it quantitative} but
indirect way through the relation between the mid and  far-infrared luminosities.
Dust enshrouded young stars emit radiation which 
is reprocessed and emitted by dust mainly in the wavelength domain 
40--1000~\micron. Hence the FIR luminosity is
one of the best tracers of the global SFR \citep{Kennicutt98a}.
In a sample of nearby IRAS galaxies observed with ISOCAM, \citet{Chary01} found
strong correlations
between the luminosity at 12 and 15~\micron\ and the total IR luminosity at $\lambda$8--1000~\micron.
This result, which confirms the early works based on IRAS studies \citep[e.g.][]{Spinoglio95},
 corroborates the idea that the MIR regime is a good tracer of the star formation rate
in galaxies. 
At the redshift of Abell 1689, the central wavelength of the LW3 filter corresponds
to rest--frame $\lambda=$ 12.7~\micron, which is close to the central wavelength of the 
IRAS 12~\micron\ band. We converted the LW3 fluxes into IRAS equivalent 12~\micron\
fluxes, taking into account the differences in the bandwidth and response function
between the two filters.
Adopting the MIR--FIR relations derived by  \citet{Chary01},
we estimated the far-infrared luminosity for all LW3 sources in A1689.
For LW3--detected galaxies, we find that
the infrared luminosity ranges between $0.9\x 10^{10}~\Lo$
and $6.2\x 10^{10}~\Lo$\footnote{Note that below infrared luminosities of 
$10^{10}~\Lo$, the contamination by cirrus might
be important. In this sample, it is negligible.}. 
Therefore, none of the MIR sources in A1689 reaches the luminosities
of Luminous Infrared Galaxies.
The average infrared luminosity is $2\x 10^{10}~\Lo$.

We computed the star formation rate from the IR luminosity following the conversion of
\citet{Kennicutt98a}:

\[ SFR(IR) = 1.7\x 10^{-10}~(L_{\rm ir}/\Lo)~ \usfr \]

The assumptions are continuous bursts lasting 10--100 Myr and the same Salpeter IMF as that 
 adopted when deriving $SFR([OII])$.
With this conversion, given the sensitivity of our ISOCAM survey,
galaxies with a SFR lower than 1.4~\usfr\ would not have been detected.

The SFR(IR)s derived in this way are listed in Table~\ref{tab:sfr}.
In Abell 1689, the LW3 sources with accurate LW3 flux (11 in 16) 
have a SFR(IR) that ranges between 1.6 and 11 \usfr\
with an average of 3.5~\usfr\ per galaxy and a median value of 2.3 \usfr.

Deep
VLA radio maps of Abell 1689 will soon become available
\citep{Morrison00}.  The radio continuum emission provides another
excellent tool to probe the total star formation rate in
galaxies, therefore it will be interesting to compare
these new data with our MIR-based
estimate of the star formation rate.

Comparisons with the coeval field are still difficult due to the lack of data.
 The situation will remain as such until  the ELAIS ISOCAM surveys
are fully released \citep{Oliver00}. For the time being, 15~$\micron$ data have been published for a couple of
small fields: the CFRS field 1415+52 \citep{Flores99b} and the Hubble Deep field
\citep{Serjeant97,Aussel99} which includes galaxies at an   average redshift of  $z=0.8$.  At this distance, given their
 completeness limits  (0.25 mJy for the CFRS 
field and 0.10 mJy for the HDF), these surveys picked up  mostly luminous infrared galaxies
($\Lir > 10^{11} \Lo$) with star formation rates much higher than in Abell 1689.
The ISOCAM survey of the whole CFRS fields provided a sample of 5 galaxies with a redshift
between 0.1 and 0.3 and a 15~\micron\ flux higher than the sensitivity limit of the A1689 survey 
(Flores et al., in preparation). Their median SFR inferred from their mid--infrared
flux with the same method as in  this paper, 2.2~\usfr, is similar to the median SFR measured
in the star--forming galaxies of Abell 1689.
A comparison per morphological type, which would be more fair, is yet impossible  due to  the
size of these samples.

\subsection{Hidden star formation in Abell 1689}
We will now compare 
the star formation rates derived in the infrared and in the optical for the
LW3--detected galaxies.

\subsubsection{Star formation missed in the optical}
\begin{figure}
\centerline{\psfig{file=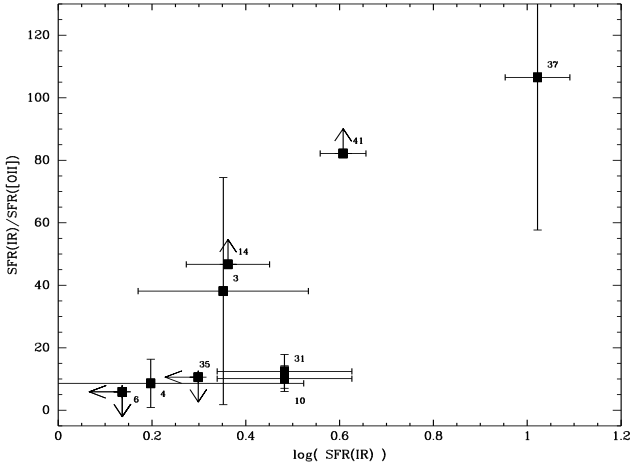,width=\columnwidth,angle=-90}}
\caption{Ratio between the star formation rate derived from the 
infrared to the star formation estimated from the flux of the
\OII\ line versus log SFR(IR) (in \usfr).  No extinction ot aperture 
corrections have been applied to SFR(\OIIt). The ISO ID number is indicated.} 
\label{fig:sfr}
\end{figure}

As shown in Figure~\ref{fig:sfr}, 
the ratio SFR(IR)/SFR([OII]) ranges between about 10 and 100  with a median value of 12  
(a mean of 30)
when only considering emission-line galaxies for which the \OIIt\ flux could be measured
and for which the 15~\micron\ flux is well determined (5 galaxies in total).
 Adding the LW3 sources with undetected \OII\ line and assigning to its flux a conservative
 upper limit of $5 \x 10^{-17}~\uflux$, the median SFR(IR)/SFR(\OIIt) raises to 30.
Therefore, one misses at least 90\% of the star formation
activity when estimating the latter from the \OIIt\ optical line. 

 Note that
a high SFR(IR)/SFR(\OIIt) is more likely the result of a high extinction rather than
a strong underestimate of the \OIIt\ flux caused by slit aperture effects. Surely, 
   excentered \HII\ regions might have been missed by the slitlets which
 were centered on the nucleus and had a random orientation with respect to the
galaxy position angle.
We tried to quantify aperture losses extrapolating the surface brightness
  profile of the  \OIIt\ line along our 1.3''--wide slits to the 
entire galaxy. We obtained upper limits for the  aperture correction factor 
 ranging between  1.2 and 1.9 with an average of 1.6.
Therefore, even after applying this correction, SFR(IR) and SFR(\OIIt) cannot be
reconciled without a strong extinction.
We remind the reader that the ``standard'' dust correction generally adopted
for spirals (1 mag at $\rm H\alpha$) is equivalent,  given the SFR calibrations
of \citet{Kennicutt92},  to assuming SFR(IR)/SFR(\OIIt)=2.5 .  
Undoubtedly, an important fraction of the star formation activity in Abell 1689 is dust enshrouded.
Using deep radio centimetric observations to trace star formation in the $z=0.41$  cluster
CL0939+4713, \citet{Smail99} reached a similar conclusion.

The level of dust enshrouded star formation depends however very much on the galaxy spectral type,
as can be seen in Figures~\ref{fig:lw3vlw2}b and \ref{fig:lw23vlw3k}b where the average LW3 luminosity,
 LW3/K' and LW3/LW2 flux ratios are plotted for each spectral class. 

In Section~3.3.4, it has been reported that all e(a) galaxies in our sample
are LW3 sources and that about 25\% of the LW3 sources have e(a) spectra.
We see from Figure~\ref{fig:lw3vlw2}b that
galaxies with an e(a) signature have the largest LW3 luminosity and LW3/LW2 flux ratio, i.e. their MIR
emission is clearly  powered by a starburst rather than quiescent star--formation. In the most extreme
galaxy among these dusty starbursts, ISO\#37, the total SFR is underestimated by a factor of 100 when
 computed from the [OII] line (see Fig.~\ref{fig:sfr}).
We note however that, whereas e(a)'s are common among the LW3 subsample and have rather large SFRs, they
 constitute in number only 5\% of the total cluster population.

It was suggested that some of the numerous k+a galaxies could be extreme e(a)'s in which the
[OII] line is totally extinguished \citep{Smail99}.
 However, no k+a is detected in the mid--infrared regime in
Abell 1689. Galaxies
of this spectral class do not host hidden star formation activity up to the ISOCAM sensitivity limit of
 1.4~\usfr. They are rather  likely to be  genuine ``post-starburst''
galaxies or at least galaxies where star formation has stopped.

It has been noted in Section~3.3.4 that some of the LW3 sources do not show
signs of SF activity in their spectra. In fact,
at least five galaxies \footnote{Out of a total of 16 LW3 sources (30\%).
 The maximum percentage is 50\% taking into account the three  
 LW3 sources absent in our MOS sample.}, namely ISO\#3,14,26,30,32 
that are considered as 'passive' in
 view of their optical
properties -- a k-/k(e)-type spectrum and, sometimes, an early--type morphology (Es,S0s)   -- are
detected in the 15~\micron\ band. 
The emission at the same rest-frame wavelength from ellipticals in Virgo and Coma
was interpreted as  dominated by photospheric emission from old
 stars \citep{Boselli98}. However, stellar emission is not likely to be
the dominant source of the 15 micron flux observed in the optically-passive galaxies of A1689,
as judged from their LW3/LW2 and LW3/K' flux ratios (discussed in Sec.~\ref{sec:irsfr}).
They are much higher than those of Virgo's ellipticals while the 15~\micron\ luminosity
increases by  a factor of 5  (see Fig.~\ref{fig:lw3vlw2}b and \ref{fig:lw23vlw3k}b).
Placed at the redshift of Abell 1689,
 the nearby ellipticals from the sample of \citet{Boselli98} would have LW3 fluxes
below 0.1 mJy; therefore none of them would have been detected by our survey.
If part of the LW3 emission of the 'passive' A1689 galaxies is really, as we believe, due to star
formation, the inferred value of the SFR is modest but significant: up to 2 \usfr\ 
(1 \usfr\ applying a correction factor due the photospheric emission) and
20--40 times higher than that derived in the optical from  the upper limit of the \OIIt\ line
 (see Fig.~\ref{fig:sfr}).

\subsubsection{Star formation missed in the infrared}
Conversely, one could ask whether all star formation shows up in the mid-infrared.
In principle, in environments with
 small amounts of dust, the UV photons emitted in SF regions are not re-processed and are
 missed in the MIR.
In our sample, one galaxy with strong equivalent widths of the optical emission lines was not detected at 15~\micron (MOS\#48).
It turns out however that it has an R--band magnitude below 19.5. No galaxies have been detected
in LW3  in this luminosity domain, apart from one object with R=20.4 (ISO\#39) for which the
cluster membership is uncertain as it is based on its photometric redshift.
The 'optical' SFR measured for  galaxy  MOS\#48  is only 0.15~\usfr.
The non-detection is then likely due to the lack of sensitivity of ISOCAM for low mass galaxies with
integrated SFR below 1.4~\usfr. 
Therefore, this galaxy would not have been detected even if the infrared to optical SFR was a factor of 10.
Sensitivity problems on the MIR side may also explain why two e(c)
galaxies in the ISOCAM field of view were not detected as well (MOS\#2 and MOS\#69). 

 Given the low-level of star formation undetected by ISOCAM in faint
blue galaxies and the paucity of this kind of objects even in the
photometric sample (see the color-magnitude diagram in Fig.~\ref{fig:CMRlw}),
the MIR emission should provide a 
good rough estimate of the {\em total} SFR in the cluster.
Of course, our optical and infrared surveys  missed the  faint end population of
 star--forming dwarf galaxies. If their luminosity function is steep enough, one 
might underestimate  significantly  the total star formation rate in the cluster.
 
\subsubsection{Dust extinction and the cluster environment}
\label{sec:env}

 The high 15~\micron\ luminosities, LW3/K' color index and SFR(IR)/SFR(opt) ratios
 measured in A1689
clearly indicate that the star-forming galaxies in the cluster
are affected by strong dust obscuration.
Does this high level of extinction result from the cluster environment ?  
Such an effect would not be unexpected. Ram-pressure and galactic harassment
contribute to strip the outer  gas reservoir of gas--rich galaxies falling in the cluster
environment. The star--formation activity is then quenched in the disk while 
it is maintained longer in the central regions which are less under the influence of the outside
 environment but are more affected by internal extinction. 
The increasing contribution of the dust--obscured nuclear or circumnuclear SF would be
at the origin of a correlation between the apparent  extinction  and the \HI\ deficiency.
Indeed, studying a sample of nearby spiral galaxies mapped by ISOCAM, 
\citet{Roussel01} found that the ratio of the MIR to optical diameter decreases when
the \HI\ deficiency increases while the mean FIR to \Ha\ flux ratio rises along with central
 concentration of the MIR emission and hence of the warm dust.
Besides, according to \citet{Moss00}, circumnuclear star--formation is more common
in rich clusters than in the field.

 On the other side, 
dust enshrouded star formation episodes are commonly found in field luminous
 infrared galaxies \citep{Sanders96b} -- a lot of them being however interacting systems --,
 and recent studies indicate that most of the star--formation activity could be
 hidden even in UV--bright low--mass galaxies, such as the blue compact dwarf
galaxies \citep{Hunt01}.  It is then not unlikely that SF in some, 
if not the majority,
of the galaxies that fell in the cluster was already dust obscured.
Understanding whether dust extinction is enhanced in galaxies as a consequence
of the cluster environment will be possible only when
MIR field surveys will have provided sufficient
data for a statistical comparison with clusters such as A1689.

%-----------------
\section{Summary and conclusions}
%-----------------
We have presented the follow-up study of a mid--infrared ISOCAM survey
of the $z=0.18$ rich cluster of galaxies Abell 1689 (Paper I).  We
compiled optical imaging, photometric and spectroscopic data for about
100 galaxies lying within the inner  $\rm Mpc^2$  of the
cluster. This database was used to assess their cluster membership 
 likelihood and
characterize their morphological type and spectral class.  The
spectroscopic sample includes 75\% of the galaxies  brighter than 
$R=17.75$  with a photometric
redshift consistent with a cluster membership and 75\% of the ISOCAM
mid--infrared members. The optical properties of the optically
selected  cluster members seem to already differ from those
of  the galaxy populations in 
typical rich clusters in the nearby universe. In particular, we
confirm the high fraction of blue galaxies previously reported in this
cluster. Focussing on the properties of the mid--infrared emitters,
we found that:\\
 (1) The redshift histogram of ISOCAM galaxies towards A1689
shows   a prominent excess of sources at the  baricentric
redshift of the cluster, $z=0.184$ as determined from our
spectroscopic sample. They are undoubtly related to the cluster
environment. \\   
 (2)  The majority of the  sources detected at 15~\micron\   are
 luminous, blue, emission--line,  morphologically  disturbed  spiral galaxies, 
i.e. the star--forming galaxies usually associated with the 'Butcher--Oemler'
 effect. Sources with the lowest LW3/LW2 (15~\micron\ to 6.75~\micron)
flux ratio, in particular those undetected at 15~\micron,
 typically consist of luminous early--type galaxies  with
passive spectra, i.e. with no signs  of current or recent star
 formation.\\
(3) However, at least  five 15~\micron\ sources  ($\geq 30$\% of  the LW3--subsample)
 do not show
any sign of star--formation activity in their optical spectrum.\\
(4) More than 70\% of the  emission--line galaxies in our
spectroscopic sample are detected at 15~\micron. The star--forming
dwarf galaxies were too faint to be detected by ISOCAM.
  All four cluster members 
with weak to moderate emission lines and unusually
strong early Balmer lines in absorption 
('e(a)' class), typical of dusty starbursts, are LW3 sources. They have
the highest  15~\micron\  to 6.75~\micron\ flux ratio in our sample.
 None of the galaxies
with a post-starburst optical spectrum ('k+a' galaxies)
have been detected at 15\micron. \\

The relative high values of the LW3/LW2 and LW3/K' color indexes 
 indicate that the LW3 luminosity of the 15\micron\ emitters is
not dominated by stellar emission. The AGN activity is at low level
in the cluster.
 The ISOCAM LW3 emission is hence a reliable tracer of
the dust--obscured star formation activity.  We have then estimated
the total infrared luminosity and inferred star formation rate from
the 15~\micron\ (12.7~\micron\ rest--frame) luminosity and compared it
 with that derived in the
optical from the flux of the \OIIt\ line. We found that:\\
 (1) There is no instance of Luminous Infrared Galaxies (LIGs) in Abell 1689.
The highest total infrared luminosity, $6.2 \x 10^{10}~\Lo$, is
measured in an e(a) galaxy with a derived star formation rate of
about 11~\usfr. Its high LW3/LW2 ratio is consistent with 
star formation occurring in a  dust-enshrouded burst mode. SFRs up to
 2~\usfr\ are estimated  in the apparently passive k--type galaxies. As seen from
 the MIR window, these objects form stars in a mode typical of spirals.
On the other hand, the upper limit of the total SFR for k+a galaxies, 1.4~\usfr, is
consistent with the absence of a strong SF activity in these objects.\\ 
 (2) The median SFR(IR) of the LW3--detected
galaxies is 2~\usfr\ while the median SFR(opt) of the \OIIt--detected
galaxies is  only 0.2~\usfr. For galaxies in common to both samples, the ratio
 SFR(IR)/SFR(opt) is very high and ranges between 10 and 100,
being highest among e(a) galaxies.  An underestimate of the \OIIt\
luminosity and hence of SFR(opt) due to slit effects alone cannot
account for the differences between the optical and infrared 
indicators of the star formation rate. \\

We conclude that a significant  portion of the star formation activity
taking place in Abell 1689 is hidden in the cluster members when
observed through the optical window. At least 90\% of the SF
 is missed when estimated  from the \OIIt\ line,  which is known
anyway as an unreliable, although often used,  tracer of the SFR.  

Comparing the SFR and the amount
of dust extinction of A1689 members with those of coeval field galaxies
would be useful to investigate in details the effects of the cluster environment
on the star formation history of galaxies.  It would however require to
understand the complex dynamics of this rich cluster and, in particular,
to locate the MIR sources in its different sub-units.

\begin{acknowledgements}
First of all, we wish to thank members of the MORPHS collaboration, in particular
Warrick Couch and Ian Smail 
for providing us the reduced HST images and morphological info 
 prior to publication. The input we got from Ian Smail on the original
manuscript helped a lot improving it. We thank the referee, Dr. J{\o}rgensen,
for his useful comments. 
 We are grateful to Chris Lidman, Adam Stanford, 
 Roberto De Propris and Simon Dye for giving us their optical/near--infrared images and
photometric data towards Abell 1689. Many thanks to the NTT team at La Silla for their
precious help with the spectroscopic observations. PAD and BMP  acknowledge support from
 the network Formation and Evolution of Galaxies set up by the European Commission under
 contract ERB FMRX CT96086 of its TMR program. 
\end{acknowledgements}

%--------------------------------------------------------------------
%           Bibliography  (generated by bibtex)
%--------------------------------------------------------------------

\bibliographystyle{ms1971}
\bibliography{ms1971}

\begin{thebibliography}{65}
\expandafter\ifx\csname natexlab\endcsname\relax\def\natexlab#1{#1}\fi
\expandafter\ifx\csname url\endcsname\relax
  \def\url#1{\texttt{#1}}\fi
\expandafter\ifx\csname urlprefix\endcsname\relax\def\urlprefix{URL }\fi

\bibitem[{{Abraham} et~al.(1996){Abraham}, {Smecker-Hane}, {Hutchings},
  et~al.}]{Abraham96}
{Abraham}, R.~G., {Smecker-Hane}, T.~A., {Hutchings}, J.~B., et~al., 1996, \apj
  471, 694

\bibitem[{{Aussel} et~al.(1999){Aussel}, {Cesarsky}, {Elbaz}, \&
  {Starck}}]{Aussel99}
{Aussel}, H., {Cesarsky}, C.~J., {Elbaz}, D., {Starck}, J.~L., 1999, \aap 342,
  313

\bibitem[{{Balogh} \& {Morris}(2000)}]{Balogh00b}
{Balogh}, M.~L., {Morris}, S.~L., 2000, \mnras 318, 703

\bibitem[{{Balogh} et~al.(1999){Balogh}, {Morris}, {Yee}, et~al.}]{Balogh99}
{Balogh}, M.~L., {Morris}, S.~L., {Yee}, H. K.~C., et~al., 1999, \apj 527, 54

\bibitem[{{Balogh} et~al.(2000){Balogh}, {Navarro}, \& {Morris}}]{Balogh00a}
{Balogh}, M.~L., {Navarro}, J.~F., {Morris}, S.~L., 2000, \apj 540, 113

\bibitem[{{Barger} et~al.(1998){Barger}, {Aragon-Salamanca}, {Smail},
  et~al.}]{Barger98}
{Barger}, A.~J., {Aragon-Salamanca}, A., {Smail}, I., et~al., 1998, \apj 501,
  522

\bibitem[{{Bertin} \& {Arnouts}(1996)}]{Bertin96}
{Bertin}, E., {Arnouts}, S., 1996, \aaps 117, 393

\bibitem[{{Biviano} et~al.(1997){Biviano}, {Katgert}, {Mazure},
  et~al.}]{Biviano97}
{Biviano}, A., {Katgert}, P., {Mazure}, A., et~al., 1997, \aap 321, 84

\bibitem[{{Boselli} et~al.(1998){Boselli}, {Lequeux}, {Sauvage},
  et~al.}]{Boselli98}
{Boselli}, A., {Lequeux}, J., {Sauvage}, M., et~al., 1998, \aap 335, 53

\bibitem[{{Butcher} \& {Oemler}(1984)}]{Butcher84}
{Butcher}, H., {Oemler}, A., 1984, \apj 285, 426

\bibitem[{{Cesarsky} et~al.(1996){Cesarsky}, {Abergel}, {Agnese},
  et~al.}]{Cesarsky96}
{Cesarsky}, C.~J., {Abergel}, A., {Agnese}, P., et~al., 1996, \aap 315, L32

\bibitem[{{Charlot} \& {Longhetti}(2001)}]{Charlot01}
{Charlot}, S.~., {Longhetti}, M., 2001, \mnras 323, 887

\bibitem[{{Chary} \& {Elbaz}(2001)}]{Chary01}
{Chary}, R., {Elbaz}, D., 2001, \apj 556, 562

\bibitem[{{Couch} et~al.(2001){Couch}, {Balogh}, {Bower}, et~al.}]{Couch01}
{Couch}, W.~J., {Balogh}, M.~L., {Bower}, R.~G., et~al., 2001, \apj 549, 820

\bibitem[{{Couch} \& {Sharples}(1987)}]{Couch87}
{Couch}, W.~J., {Sharples}, R.~M., 1987, \mnras 229, 423

\bibitem[{{de Propris} et~al.(1999){de Propris}, {Stanford}, {Eisenhardt},
  et~al.}]{DePropris99}
{de Propris}, R., {Stanford}, S.~A., {Eisenhardt}, P.~R., et~al., 1999, \aj
  118, 719

\bibitem[{{Dressler}(1987)}]{Dressler87}
{Dressler}, A., 1987, in: Nearly Normal Galaxies. From the Planck Time to the
  Present, Springer-Verlag, New York,  276

\bibitem[{{Dressler} et~al.(1999){Dressler}, {Smail}, {Poggianti},
  et~al.}]{Dressler99}
{Dressler}, A., {Smail}, I., {Poggianti}, B.~M., et~al., 1999, \apjs 122, 51

\bibitem[{{Dye} et~al.(2001){Dye}, {Taylor}, {Thommes}, et~al.}]{Dye01}
{Dye}, S., {Taylor}, A.~N., {Thommes}, E.~M., et~al., 2001, \mnras 321, 685

\bibitem[{Elbaz et~al.(2001)Elbaz, Cesarsky, Chanial, et~al.}]{Elbaz01}
Elbaz, D., Cesarsky, C., Chanial, P., et~al., 2001, \aap submitted

\bibitem[{{Fabricant} et~al.(1991){Fabricant}, {McClintock}, \&
  {Bautz}}]{Fabricant91}
{Fabricant}, D.~G., {McClintock}, J.~E., {Bautz}, M.~W., 1991, \apj 381, 33

\bibitem[{{Fadda} et~al.(2000{\natexlab{a}}){Fadda}, {Elbaz}, {Duc}, \&
  {Flores}}]{Fadda00b}
{Fadda}, D., {Elbaz}, D., {Duc}, P.-A., {Flores}, H., 2000{\natexlab{a}}, in:
  ASP Conf. Ser. 200: Clustering at High Redshift, ~96

\bibitem[{{Fadda} et~al.(2000{\natexlab{b}}){Fadda}, {Elbaz}, {Duc},
  et~al.}]{Fadda00}
{Fadda}, D., {Elbaz}, D., {Duc}, P.~A., et~al., 2000{\natexlab{b}}, \aap 361,
  827 (Paper~I)

\bibitem[{{Fasano} et~al.(2000){Fasano}, {Poggianti}, {Couch},
  et~al.}]{Fasano00}
{Fasano}, G., {Poggianti}, B.~M., {Couch}, W.~J., et~al., 2000, \apj 542, 673

\bibitem[{{Fisher} et~al.(1998){Fisher}, {Fabricant}, {Franx}, \& {van
  Dokkum}}]{Fisher98}
{Fisher}, D., {Fabricant}, D., {Franx}, M., {van Dokkum}, P., 1998, \apj 498,
  195

\bibitem[{{Flores} et~al.(1999){Flores}, {Hammer}, {Thuan}, et~al.}]{Flores99b}
{Flores}, H., {Hammer}, F., {Thuan}, T.~X., et~al., 1999, \apj 517, 148

\bibitem[{{Genzel} \& {Cesarsky}(2000)}]{Genzel00}
{Genzel}, R., {Cesarsky}, C.~J., 2000, \araa 38, 761

\bibitem[{{Girardi} et~al.(1997){Girardi}, {Fadda}, {Escalera},
  et~al.}]{Girardi97}
{Girardi}, M., {Fadda}, D., {Escalera}, E., et~al., 1997, \apj 490, 56

\bibitem[{{Gudehus} \& {Hegyi}(1991)}]{Gudehus91}
{Gudehus}, D.~H., {Hegyi}, D.~J., 1991, \aj 101, 18

\bibitem[{{Hunt} et~al.(2001){Hunt}, {Vanzi}, \& {Thuan}}]{Hunt01}
{Hunt}, L.~K., {Vanzi}, L., {Thuan}, T.~X., 2001, \aap 377, 66

\bibitem[{{Jansen} et~al.(2001){Jansen}, {Franx}, \& {Fabricant}}]{Jansen01}
{Jansen}, R.~A., {Franx}, M., {Fabricant}, D., 2001, \apj 551, 825

\bibitem[{{Kennicutt}(1998)}]{Kennicutt98a}
{Kennicutt}, R.~C., J., 1998, \araa 36, 189

\bibitem[{{Kennicutt}(1992)}]{Kennicutt92}
{Kennicutt}, Robert~C., J., 1992, ApJ 388, 310

\bibitem[{{Kodama} \& {Bower}(2001)}]{Kodama01}
{Kodama}, T., {Bower}, R.~G., 2001, \mnras 321, 18

\bibitem[{{Laurent} et~al.(2000){Laurent}, {Mirabel}, {Charmandaris},
  et~al.}]{Laurent00}
{Laurent}, O., {Mirabel}, I.~F., {Charmandaris}, V., et~al., 2000, \aap 359,
  887

\bibitem[{{Lin} et~al.(1996){Lin}, {Kirshner}, {Shectman}, et~al.}]{Lin96}
{Lin}, H., {Kirshner}, R.~P., {Shectman}, S.~A., et~al., 1996, \apj 464, 60

\bibitem[{{Madau} et~al.(1996){Madau}, {Ferguson}, {Dickinson},
  et~al.}]{Madau96}
{Madau}, P., {Ferguson}, H.~C., {Dickinson}, M.~E., et~al., 1996, \mnras 283,
  1388

\bibitem[{{Margoniner} \& {de Carvalho}(2000)}]{Margoniner00}
{Margoniner}, V.~E., {de Carvalho}, R.~R., 2000, \aj 119, 1562

\bibitem[{{Margoniner} et~al.(2001){Margoniner}, {de Carvalho}, {Gal}, \&
  {Djorgovski}}]{Margoniner01}
{Margoniner}, V.~E., {de Carvalho}, R.~R., {Gal}, R.~R., {Djorgovski}, S.~G.,
  2001, \apjl 548, L143

\bibitem[{{Metevier} et~al.(2000){Metevier}, {Romer}, \& {Ulmer}}]{Metevier00}
{Metevier}, A.~J., {Romer}, A.~K., {Ulmer}, M.~P., 2000, \aj 119, 1090

\bibitem[{{Miralda-Escude} \& {Babul}(1995)}]{Miralda-Escude95}
{Miralda-Escude}, J., {Babul}, A., 1995, \apj 449, 18

\bibitem[{{Molinari} et~al.(1996){Molinari}, {Buzzoni}, \&
  {Chincarini}}]{Molinari96}
{Molinari}, E., {Buzzoni}, A., {Chincarini}, G., 1996, \aaps 119, 391

\bibitem[{{Monet}(1996)}]{Monet96}
{Monet}, D., 1996, in: \baas, vol. 188,  5404

\bibitem[{{Morrison}(2000)}]{Morrison00}
{Morrison}, G.~E., 2000, in: American Astronomical Society Meeting, vol. 197,
  5704

\bibitem[{{Moss} \& {Whittle}(2000)}]{Moss00}
{Moss}, C., {Whittle}, M., 2000, \mnras 317, 667

\bibitem[{{Oliver} et~al.(2000){Oliver}, {Rowan-Robinson}, {Alexander},
  et~al.}]{Oliver00}
{Oliver}, S., {Rowan-Robinson}, M., {Alexander}, D.~M., et~al., 2000, \mnras
  316, 749

\bibitem[{{Pickles} \& {van der Kruit}(1991)}]{Pickles91}
{Pickles}, A.~J., {van der Kruit}, P.~C., 1991, \aaps 91, 1

\bibitem[{{Poggianti} et~al.(2001){Poggianti}, {Bressan}, \&
  {Franceschini}}]{Poggianti01}
{Poggianti}, B.~M., {Bressan}, A., {Franceschini}, A., 2001, \apj 550, 195

\bibitem[{{Poggianti} et~al.(1999){Poggianti}, {Smail}, {Dressler},
  et~al.}]{Poggianti99}
{Poggianti}, B.~M., {Smail}, I., {Dressler}, A., et~al., 1999, \apj 518, 576

\bibitem[{{Poggianti} \& {Wu}(2000)}]{Poggianti00}
{Poggianti}, B.~M., {Wu}, H., 2000, \apj 529, 157

\bibitem[{{Rola} et~al.(1997){Rola}, {Terlevich}, \& {Terlevich}}]{Rola97}
{Rola}, C.~S., {Terlevich}, E., {Terlevich}, R.~J., 1997, \mnras 289, 419

\bibitem[{{Roussel} et~al.(2001){Roussel}, {Sauvage}, {Vigroux},
  et~al.}]{Roussel01}
{Roussel}, H., {Sauvage}, M., {Vigroux}, L., et~al., 2001, \aap 372, 406

\bibitem[{Sanders \& Mirabel(1996)}]{Sanders96b}
Sanders, D.~B., Mirabel, I.~F., 1996, ARA\&A 34, 749

\bibitem[{Sanders et~al.(1988)Sanders, Soifer, Elias, et~al.}]{Sanders88a}
Sanders, D.~B., Soifer, B.~T., Elias, J.~H., et~al., 1988, ApJ 325, 74

\bibitem[{{Serjeant} et~al.(1997){Serjeant}, {Eaton}, {Oliver},
  et~al.}]{Serjeant97}
{Serjeant}, S. B.~G., {Eaton}, N., {Oliver}, S.~J., et~al., 1997, \mnras 289,
  457

\bibitem[{{Smail} et~al.(1997){Smail}, {Dressler}, {Couch}, et~al.}]{Smail97}
{Smail}, I., {Dressler}, A., {Couch}, W.~J., et~al., 1997, \apjs 110, 213

\bibitem[{{Smail} et~al.(1998){Smail}, {Edge}, {Ellis}, \&
  {Blandford}}]{Smail98}
{Smail}, I., {Edge}, A.~C., {Ellis}, R.~S., {Blandford}, R.~D., 1998, \mnras
  293, 124

\bibitem[{{Smail} et~al.(1999){Smail}, {Morrison}, {Gray}, et~al.}]{Smail99}
{Smail}, I., {Morrison}, G., {Gray}, M.~E., et~al., 1999, \apj 525, 609

\bibitem[{{Spinoglio} et~al.(1995){Spinoglio}, {Malkan}, {Rush},
  et~al.}]{Spinoglio95}
{Spinoglio}, L., {Malkan}, M.~A., {Rush}, B., et~al., 1995, \apj 453, 616

\bibitem[{{Teague} et~al.(1990){Teague}, {Carter}, \& {Gray}}]{Teague90}
{Teague}, P.~F., {Carter}, D., {Gray}, P.~M., 1990, \apjs 72, 715

\bibitem[{{Tran} et~al.(2001){Tran}, {Lutz}, {Genzel}, et~al.}]{Tran01}
{Tran}, Q.~D., {Lutz}, D., {Genzel}, R., et~al., 2001, \apj 552, 527

\bibitem[{{Tresse} \& {Maddox}(1998)}]{Tresse98}
{Tresse}, L., {Maddox}, S.~J., 1998, \apj 495, 691

\bibitem[{{Tyson} \& {Fischer}(1995)}]{Tyson95}
{Tyson}, J.~A., {Fischer}, P., 1995, \apjl 446, L55

\bibitem[{{Veilleux} et~al.(1999){Veilleux}, {Kim}, \& {Sanders}}]{Veilleux99}
{Veilleux}, S., {Kim}, D.~., {Sanders}, D.~B., 1999, \apj 522, 113

\bibitem[{Veilleux \& Osterbrock(1987)}]{Veilleux87}
Veilleux, S., Osterbrock, D.~E., 1987, ApJS 63, 295

\end{thebibliography}

\clearpage 

\appendix

\section{Completeness of the MOS sample}
We  used the sample of photometric members by \citet{Dye01} to assess the completeness of our
MOS survey. Dye et al's survey is complete up to $B=23.7$; 
unfortunately it does not cover totally our MOS field of view (see Fig.~\ref{fig:idvel}). 
We restricted therefore our analysis to the $5'\times 6'$ field of view where 
photometric redshifts are available.

First of all, we note that the photometric redshifts are reliable:
the cluster membership of almost 90\% of the photometric members with MOS data is confirmed with
our direct redshift measurements and the average photometric and spectrophotometric redshifts
differ by less than 10\% (see Fig.~\ref{fig:histz}).

\begin{figure}
\centerline{\psfig{file=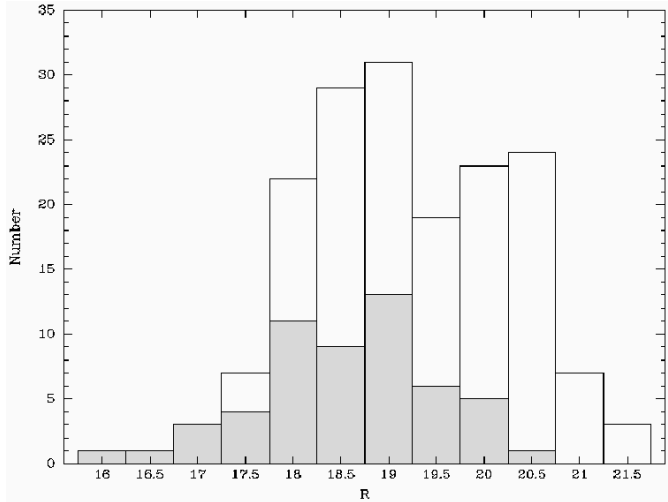,width=\columnwidth}}
\caption{R--band magnitude histogram of photometrically  and
spectroscopically (shaded) confirmed  cluster members within the area
 covered by the photometric redshift survey of Dye et al. (2001).}
\label{fig:histRcomp}
\end{figure}

\begin{figure}
\centerline{\psfig{file=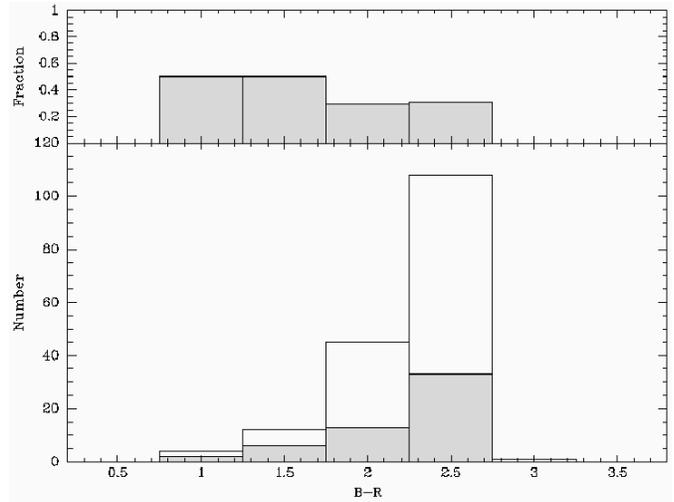,width=\columnwidth}}
\caption{$B-R$ color index histogram of photometrically and spectroscopically
 (shaded) cluster members within the area
 covered by the photometric redshift survey of Dye et al. (2001).
 The top panel shows the  number ratio of spectroscopically
to photometrically confirmed members within each color bin.}
\label{fig:histBRcomp}
\end{figure}

As shown in Figure~\ref{fig:histRcomp}, we have obtained spectra for
 75\% of the cluster members brighter than $R=17.75$. For $R<19.5$, we
 sampled only about 40\% of the photometrically confirmed cluster
members while we missed most of the faintest dwarf galaxies with $R>20$.
The histogram of the B-R color--index of photometric and spectroscopic cluster members
is presented in Figure~\ref{fig:histBRcomp}.
As shown in the top panel, the comparison of these two distributions reveals
that there is a weak color bias in our
spectroscopic sample, which includes relatively more blue than red galaxies.
Galaxies bluer than B-R $<$ 1.75 (which account for 15\% of the spectroscopic sample)
are over-represented by 60\%.
Note however that the
photometric sample  most likely misses the faintest bluest galaxies in the cluster
 because the photometric redshift technique does not work well for such deviant objects,
 and this acts so as to counterbalance the bias.

\clearpage

\section{Blue fraction in Abell 1689}
Our spectroscopic survey of Abell 1689 allows us to plot a 
color-magnitude diagram only including confirmed cluster members. One may then, in principle, 
  estimate the blue fraction without the uncertainty introduced in photometric methods by
 the background  correction. 
For a fair comparison with the latest studies on the subject,  we computed \fb\ using
 the definition  proposed by  \citet[][hereafter MdC01]{Margoniner01}. We selected all
 galaxies inside a clustercentric radius of 0.7 Mpc
\footnote{With the cosmology adopted by \citet{Margoniner01}, 0.7 Mpc
 corresponds to 3', i.e. within our field of view.} with an absolute
 red magnitude between $M^{*}-1$ and $M^{*}+2$  \citep[with M*=-20.91, ][]{Lin96} and with
 photometric data available in the catalog of MdC01. 
  Using as a reference for the blue excess fraction ($\fb=N_{\mbox{\tiny B}}/N_{\mbox{\tiny tot}}$)
 the same g-r vs r color-magnitude relation as in MdC01,
 we derived 
$\fb=0.14\pm0.04~(N_{\mbox{\tiny tot}}=83)$ when the photometric members
are  included and 
$\fb=0.16\pm0.07~(N_{\mbox{\tiny tot}}=32)$  for the sample restricted to spectroscopically
 confirmed members.  Correcting for the  color bias discussed in Appendix~A, we find that
a lower limit to the blue fraction   is $\fb=0.14$. 

We also defined a new color-magnitude relation (CMR)
based on the B and V bands and restricted
to morphologically confirmed early-type galaxies. The B--V color index was that
 originally used in BO84.
As in BO84, the blue galaxies were defined as those with a B-V  lower by 0.2 mag
 with respect to the CMR. Using the same luminosity range and spatial domain
as before, we found $\fb=0.18\pm0.04~(N_{\mbox{\tiny tot}}=127)$ and $0.25\pm0.06~(N_{\mbox{\tiny tot}}=68)$
 for the photometric and spectroscopic cluster sample respectively
 ($\fb=0.19$)  if the latter is corrected for the color bias).

Our results arguably favor a  rather large blue fraction for this cluster --
at least 0.15 -- which is free from uncertainties in the background
subtraction.

%------------------

\end{document}